\documentclass [10pt]{iopart}
\usepackage[dvips]{graphicx}
\begin{document}
\jl{4}

\title{Deterministic approach to microscopic three-phase traffic theory}

\author{Boris S. Kerner $^1$ and Sergey L. Klenov $^2$}      


\address{$^1$
DaimlerChrysler AG, REI/VF, HPC:  G021, 71059 Sindelfingen, Germany 
}

\address{$^2$
Moscow Institute of Physics and Technology, Department of Physics, 141700 Dolgoprudny,
Moscow Region, Russia
}


\pacs{89.40.+k, 47.54.+r, 64.60.Cn, 64.60.Lx}

\begin{abstract}
  Two different 
 deterministic microscopic traffic flow models, which are in the context of the Kerner's there-phase traffic theory,
  are introduced.
  In an acceleration time delay  model (ATD-model),
 different time delays in driver acceleration associated with  driver behaviour in various
 local driving situations are explicitly incorporated into the model. Vehicle acceleration depends on local traffic situation, i.e.,
  whether a driver is within the free flow, or synchronized flow, or else wide moving jam traffic phase.
 In a speed adaptation model (SA-model), 
  vehicle speed adaptation occurs in synchronized flow   depending on
 driving conditions. It is found that
the ATD- and SA-models show spatiotemporal congested traffic patterns that are adequate with
empirical results. 
In the ATD- and SA-models, the onset of congestion in free flow at a freeway bottleneck is associated with a first-order
phase transition from free flow to synchronized flow;
moving jams emerge spontaneously in synchronized flow only.
Differences between the ATD- and SA-models are studied.
A comparison of the ATD- and SA-models with stochastic models in the context of three phase traffic theory is made. 
A critical discussion of earlier traffic flow theories and models based on the fundamental diagram approach
is presented.
\end{abstract}

\maketitle

\section{Introduction}
\label{Introduction}

Theoretical studies of freeway traffic flow dynamics is one of the rapid developing fields of statistical and nonlinear physics
(see the reviews~\cite{Gartner,Wolf,Sch,Helbing2001,Nagatani_R,Nagel2003A}, the book~\cite{KernerBook}, and the conference proceedings~\cite{Lesort,Ceder,Taylor,SW1,SW2,SW3,SW4,SW5}).
For a mathematical description of freeway traffic flow, a huge number of
 different microscopic and macroscopic    traffic flow models
have been introduced. In  macroscopic models, 
  individual dynamic vehicle behaviour is
 averaged,
i.e., these models describe dynamics of average  traffic flow characteristics like 
average vehicle speed and density (see 
e.g.,~\cite{Payne,Kuehne,KK1993,Klar,Bellomo})\footnote{It should be noted that
transferring the information delivered from one 
vehicle interacting with the neighbour ones requires to deal carefully with a complex averaging process
by derivation of   a macroscopic traffic flow model. The related mathematical theory  is developed in Ref.~\cite{Darbha1,Darbha2}.}. 
Microscopic traffic flow models describe individual dynamic vehicle behaviour, which should simulate
empirical spatiotemporal features of phase transitions and congested patterns in freeway traffic. 
In this article, we restrict a consideration of {\it microscopic} traffic flow models {\it only}.

There are two types of   microscopic traffic flow models: Deterministic
models and stochastic 
models~\cite{Gartner,Wolf,Sch,Helbing2001,Nagatani_R,Nagel2003A,KernerBook}.
In deterministic models, some dynamic rules of vehicle motion in traffic flow are responsible for 
  spatiotemporal features of traffic patterns that the models exhibit. 
Contrastingly,   stochastic models, in addition to dynamic rules of vehicle motion,
exhibit model fluctuations, which play a fundamental role for traffic  pattern features.  

There are at least two classes of deterministic traffic flow models~\cite{Gartner,Sch,Helbing2001,Nagatani_R,Nagel2003A}.
In the first class, driver time delays in vehicle acceleration (deceleration) $a$ 
are explicitly taken into account. An example is the classic model of  Herman,  Montroll,   Potts,  and
Rothery~\cite{GH1959}: If
the vehicle speed $v$, or the speed difference between the vehicle speed  and the speed of the preceding vehicle $v_{\ell}$,
  or else   the net distance $g$ (space gap) between vehicles  changes,
  then the driver accelerates (decelerates) with a time delay $\tau$~\cite{GH1959}:
\begin{equation}
a(t+ \tau)=f(v(t),v_{\ell}(t),g(t)).
\label{Eq_Herman}
\end{equation}
Based on (\ref{Eq_Herman}), Gazis, Herman, and Rothery~\cite{GH} have developed a microscopic traffic flow model,
which is capable of describing traffic beyond of instabilities;
  steady state   solutions  of this model lie
on a one-dimensional curve in the flow--density plane (the fundamental diagram)
(see the review
by Nagel et al.~\cite{Nagel2003A} for more detail). Recall that steady state solutions are hypothetical
model solutions in which
all vehicles move at the same time-independent speed and the same space gap  between vehicles.
One of  the  mathematical descriptions of this model class first  proposed by
Nagatani and Nakanishi~\cite{NagataniNak1998} and further developed by Lubashevky et al.~\cite{Lub2003A} reads as follows
\begin{equation}
\frac{da}{dt}=\frac{f(v(t),v_{\ell}(t),g(t))-a(t)}{\tau}.
\label{Eq_a_opt}
\end{equation}
In  both models~\cite{GH,NagataniNak1998,Lub2003A}, steady state model solutions in the flow--density plane lie
on the fundamental diagram.

There is also another class of deterministic microscopic models in which
the vehicle speed satisfies the equation~\cite{Gartner,Sch,Helbing2001,Nagatani_R,Nagel2003A}:
\begin{equation}
\frac{dv}{dt}=\phi(v(t),v_{\ell}(t),g(t)).
\label{Eq_v_opt}
\end{equation}
Examples  are optimal velocity (OV) models  of Newell~\cite{New2},  Whitham~\cite{Wh},
 Bando, Sugiyama et al.~\cite{B1995A}, and the intelligent driver model (IDM) of Treiber and Helbing~\cite{Treiber1999AA,Helbing2000}.
Steady state solutions of this model class that obviously satisfy the conditions $\phi(v,v_{\ell},g)=0$ and $v=v_{\ell}$
lie on the fundamental diagram in the flow--density plane.

If  functions and model parameters in the models (\ref{Eq_a_opt}) and
 (\ref{Eq_v_opt}) 
are chosen in an 
appropriated way, then there is a range of   vehicle density in which steady state model solutions for free flow
 are unstable. This
instability, which should explain the onset of congestion, leads to wide moving jam
emergence in free flow (F$\rightarrow$J transition)~\cite{Wolf,Sch,Helbing2001,Nagatani_R,Nagel2003A}.

However, as explained in the book~\cite{KernerBook}, the above models that are
in the context of the fundamental diagram approach, as well as all other  
 traffic flow  models reviewed in~\cite{Gartner,Wolf,Sch,Helbing2001,Nagatani_R,Nagel2003A}
cannot explain   the fundamental empirical feature of traffic breakdown, i.e., that the
 onset of congestion in free flow at a bottleneck is associated with
a local first-order phase transition from free flow to synchronized flow
(F$\rightarrow$S transition)~\cite{KR1997,Kerner1998B,Kerner2002B} rather than with an
F$\rightarrow$J transition. For this reason,
 Kerner introduced a three-phase traffic theory.
  In this theory, there are three traffic phases:
(i) free flow, (ii) synchronized flow, and (iii) wide moving jam.

The first microscopic models in the context of three-phase traffic theory introduced in 2002
 are stochastic models~\cite{KKl,KKW}. As in empirical observations~\cite{Kerner1998B,Kerner2002B},
 in these models wide moving jams emerge spontaneously only in synchronized flow 
 (S$\rightarrow$J transition), i.e., the models exhibit the sequence of  F$\rightarrow$S$\rightarrow$J transitions
 leading to wide moving jam emergence in free flow;
 in addition, the models
  show all types of congested patterns found
  in empirical observations~\cite{KKl,KKW,KKl2003A,KKl2004AA,KernerBook,KKH_empirical}.
 Recently, some new  microscopic models based on three-phase traffic theory have been 
 developed~\cite{Davis2003B,Lee_Sch2004A,Jiang2004A}.
 However, there are no deterministic models in the context of three-phase traffic theory, which can exhibit
 the F$\rightarrow$S$\rightarrow$J transitions found in empirical observations and the diagram of congested
 patterns of three-phase traffic theory~\cite{Kerner2002B,KernerBook}. 
 In stochastic models~\cite{KKl,KKW,KKl2003A,KKl2004AA,KernerBook}, driver time delays in acceleration (deceleration) 
 are simulated mainly through the use of model fluctuations. Therefore, a development of   deterministic models based
  on three-phase traffic theory
 is   important for a more realistic theory   of car following behaviour.

In this paper, two deterministic microscopic three-phase traffic models are presented. In an 
{\bf a}cceleration {\bf t}ime {\bf d}elay  model
(ATD-model for short; Sect.~\ref{TimeDelayS}),   
 an explicit  description of
 driver time delays in vehicle acceleration (deceleration) is used. 
 In a {\bf s}peed {\bf a}daptation model (SA-model for short; Sect.~\ref{SA_Section}), 
 vehicle speed adaptation occurs in synchronized flow  depending on
 driving conditions. 
  In Sects.~\ref{PT_Onramp} and~\ref{SA_S},
   we show that these models exhibit the F$\rightarrow$S$\rightarrow$J transitions and congested patterns associated with
 results of empirical observations. In addition,  
 a stochastic SA-model is introduced and compared with
 the deterministic  SA-model of Sect.~\ref{SA_Section}. In Sect.~\ref{Discussion},
 the deterministic microscopic three-phase traffic models of Sects.~\ref{TimeDelayS} and~\ref{SA_Section}
 are compared with
 earlier deterministic models and a critical discussion of models in the context of the fundamental diagram approach
 is performed.

\section{Acceleration Time Delay  Model}
\label{TimeDelayS}

\subsection{Driver Behavioural Assumptions and Empirical Basis of ATD-Model \label{D_B_A}}

A deterministic three-phase traffic flow model with driver time delays (ATD-model)
is based on 
the following empirical features of phase transitions and congested patterns as well as
driver behavioural assumptions of 
three-phase traffic
theory (Sects. 2.3, 2.4, and 8.6 of the book~\cite{KernerBook}): 

(i)
In synchronized flow, a driver accepts a range
of different hypothetical steady states with various space gaps   $g$
at the same vehicle speed $v$, i.e.,
steady states of synchronized flow cover a two-dimensional region
in the flow--density plane. 

(ii) To avoid collisions, in the steady states a driver does not 
accept the vehicle speed   that is higher than some safe speed   
(denoted by $v_{\rm s}(g, \ v_{\ell})$) that
 depends on the speed of the preceding vehicle $v_{\ell}$. In contrast with earlier models in which a safe speed
 determines a multitude of steady states on the fundamental diagram~\cite{Gartner,Sch,Helbing2001,Nagatani_R,Nagel2003A,Gipps}, in
 the ATD-model the  safe speed determines
the upper boundary of the two-dimensional region for the steady states
in the flow--density plane~\cite{KernerBook}. 

(iii) If a driver cannot pass the preceding vehicle,
then the driver tends to adjust the speed to the preceding vehicle within
a synchronization gap $G(v, \ v_{\ell})$, i.e., at
\begin{equation} 
g\leq G(v, \ v_{\ell})
\label{g_G}
\end{equation}
 a speed adaptation effect
occurs.
The synchronization gap determines the lower boundary of the two-dimensional region for the steady states
in the flow--density plane. In the ATD-model, 
the speed adaptation effect is modelled  through a driver acceleration    
$K(v, \ v_{\ell})(v-v_{\ell})$ adjusting the speed to the preceding vehicle under the conditions (\ref{g_G}); 
 $K(v, \ v_{\ell})$ is a sensitivity.
 
 (iv) 
 In traffic flow with   greater space gaps, a driver searches for the opportunity
 to accelerate and to pass. This leads to driver over-acceleration, which is modelled
 through a driver acceleration    
$A(V^{\rm (free)}(g)-v)$ adjusting the vehicle speed at
 \begin{equation} 
g> G(v, \ v_{\ell})
\label{G_g}
\end{equation}
  to a   gap-dependent optimal   speed  in free flow $V^{\rm (free)}(g)$, where 
 $A$ is a sensitivity of this effect.   
 
 A competition between the speed adaptation effect
 and driver over-acceleration simulates a first-order
 F$\rightarrow$S   transition leading to the onset of congestion in real traffic flow 
 (see explanations in Sect.  2.4 in~\cite{KernerBook}).

(v) In empirical observations, due to an F$\rightarrow$S   transition there is
 a maximum point
of free flow associated with the   maximum density $\rho^{\rm (free)}_{\rm max}$, maximum flow rate $q^{\rm (free)}_{\rm max}$, and 
maximum speed
$v^{\rm (free)}_{\rm min}$ given by  the formula $v^{\rm (free)}_{\rm min}=q^{\rm (free)}_{\rm max}/\rho^{\rm (free)}_{\rm max}$
(Sect. 2.3 in~\cite{KernerBook}). This maximum point  is modelled through   F$\rightarrow$S transition, which occurs
already due to infinitesimal local perturbations in steady states of free flow
associated with 
the  optimal speed in free flow  $V^{\rm (free)}(g)$ at the density 
$\rho^{\rm (free)}_{\rm max}$. 
 
 (vi) In high density flow, a driver decelerates stronger than it is required to avoid collisions if the
 preceding vehicle begins to decelerate unexpectedly (driver over-deceleration).
 In the ATD-model, the over-deceleration effect, which explains and simulates moving jam emergence in synchronized flow,
  is modelled by
a driver  time delay $\tau$ in
 reduction of a current driver deceleration (denoted by $\tau=\tau^{\rm (dec)}_{1}(v)$). The longer $\tau^{\rm (dec)}_{1}$,
  the   stronger the 
over-deceleration effect.
In empirical observations, the lower the synchronized flow speed, the greater the probability for
moving jams emergence (Sect. 2.4 in~\cite{KernerBook}). For this reason, 
$\tau^{\rm (dec)}_{1}(v)$ is chosen to be longer at lower   speeds than at higher ones.

(vii) At the downstream front of a wide moving jam or a synchronized flow region, a driver
within the jam or the synchronized flow region does not accelerate before the preceding vehicle has begun
to accelerate. In the ATD-model, this effect is modelled through the use of
a mean driver time delay in acceleration at the downstream front of the synchronized flow region, which depends on
 a time delay in driver acceleration (denoted by $\tau=\tau^{\rm (acc)}_{0}$) and on 
 the sensitivity  $K(v, \ v_{\ell})$  at $v<v_{\ell}$. 
At the downstream front of a wide moving jam,
a mean time delay in acceleration from a standstill  $v=0$
within the jam  should be longer than the   mean driver time delay in synchronized flow~\cite{KernerBook}. 
To simulate this longer mean time delay
in vehicle acceleration, in addition with two mentioned above model effects,
a vehicle  within the jam does not accelerate before
the condition 
\begin{equation}
g\geq g^{\rm (jam)}_{\rm max}
\label{g_g_jam}
\end{equation}
 is satisfied, in which $g^{\rm (jam)}_{\rm max}$ is
 the maximum space gap within the wide moving jam phase.

(viii) Moving in synchronized flow of lower speeds,
a driver comes closer to the preceding vehicle than the synchronization gap $G$. In empirical observations,
this self-compression
of synchronized flow is called    the pinch effect (Sect. 12.2 in~\cite{KernerBook}).
In the ATD-model, the pinch effect 
 is simulated through the use of two model assumptions.
Firstly, a 
time delay  in reduction of a current driver acceleration (denoted by $\tau=\tau^{\rm (acc)}_{1}$) increases  if the speed decreases. 
Secondly, the sensitivity    $K(v, \ v_{\ell})$, which describes the speed adaptation effect  (item (iii)), is chosen
   at $v\geq v_{\ell}$ different from $K(v, \ v_{\ell})$   at $v< v_{\ell}$ (item (vii)). Specifically, $K(v, \ v_{\ell})$   at $v\geq  v_{\ell}$
is chosen to be  
 smaller at low speeds
than at higher ones. As a result, at   lower speeds   vehicles choose    smaller space gaps than
 the synchronization gap  $G$.

(ix) At the upstream front of a wide moving jam or a synchronized flow region, a driver
begins to decelerate after a time delay denoted by $\tau=\tau^{\rm (dec)}_{0}$.
This delay time should describe   realistic velocities of deceleration fronts in congested traffic patterns.

\subsection{Main Equations}

An ATD-model  reads as follows:
\begin{eqnarray}
\label{coor}
\frac{dx}{dt}=v, \\
\label{speed}
\frac{dv}{dt}=a, \\
\label{acc}
\frac{da}{dt}=\left\{
\begin{array}{ll}
(a^{\rm (free)}- a)/\tau &  \textrm{at $g > G$ and $g > g^{\rm (jam)}_{\rm max}$},  \\
(a^{\rm (syn)}-a)/\tau &  \textrm{at $g \leq G$ and $g > g^{\rm (jam)}_{\rm max}$}, \\
(a^{\rm (jam)}-a)/\tau &  \textrm{at $0\leq g \leq g^{\rm (jam)}_{\rm max}$},
\end{array} \right. 
\end{eqnarray}
where 
$x$ is  
the vehicle space co-ordinate; 
$g=x_{\ell}-x-d$;
the lower index $\ell$ marks variables related to the preceding vehicle; all vehicles have the same length $d$,
which includes the minimum space gap between vehicles within a wide moving jam; 
$a^{\rm (free)}$, $a^{\rm (syn)}$, and $a^{\rm (jam)}$ are vehicle accelerations (deceleration)
in the free flow, synchronized flow, and wide
moving jam
phases, respectively. If  the condition (\ref{G_g}) is satisfied,
 then a  vehicle moves in accordance with the rules for free 
flow.
Within synchronized flow associated with the condition $g^{\rm (jam)}_{\rm max}<g\leq G(v, \ v_{\ell})$,
the vehicle tends to adapt the speed to the preceding vehicle.
Within a wide moving jam, the space gap is small, specifically $g\leq g^{\rm (jam)}_{\rm max}$, and
the vehicle decelerates.\footnote{Since the vehicle speed $v$ cannot be negative, the following condition is also used
for Eqs. (\ref{speed}), (\ref{acc}):
\begin{eqnarray}
a(t) \geq 0 \  \textrm{at $v(t)=0$}.
\label{acc_b}
\end{eqnarray}
To satisfy this condition in numerical simulation, the acceleration $a(t)$  is replaced by the value
$\max(a(t), \ 0)$ if $v(t)=0$ at time $t$.}

\subsection{Driver Acceleration}

The accelerations (decelerations) $a^{\rm (free)}$, $a^{\rm (syn)}$, and $a^{\rm (jam)}$ are found from the condition
\begin{eqnarray}
a^{(\rm phase)}= \min( \max (\tilde a^{\rm (phase)}, \ a_{\rm min}), \ a_{\rm max}, \ a_{\rm s}),
\label{acc_lim}
\end{eqnarray}
the superscript $\lq\lq$phase" 
in (\ref{acc_lim}) means either  $\lq\lq$free", or $\lq\lq$syn", or else $\lq\lq$jam" for the related traffic 
phase;
$a_{\rm s}$ is a deceleration related to safety requirements;
$a_{\rm min}$  and $a_{\rm max}$ ($a_{\rm min}<0$, $a_{\rm max} \geq 0$) are respectively
the minimum and   maximum accelerations for cases in which there are no safety 
restrictions.
In (\ref{acc_lim}), functions  $\tilde a^{\rm (free)}$,  $\tilde a^{\rm (syn)}$, and $\tilde a^{\rm (jam)}$ 
associated with driver acceleration within the related traffic phase -- free flow, or synchronized flow, or else wide moving jam --
 are determined as follows:\footnote{In the article, large enough flow rates on the main road are considered at which
congested patterns can occur at a bottleneck. For this reason,  in (\ref{a_free})
$K$  is chosen to be independent on $g$ in the free flow phase. At 
considerably smaller
flow rates in free flow, specifically, if $g$ increases, $K$ in (\ref{a_free}) should tend towards zero when $g\gg G$.}
\begin{eqnarray}
\label{a_free}
\tilde a^{\rm (free)}(g, \ v, \ v_{\ell})=A(V^{\rm (free)}(g)-v)+  \nonumber \\
K(v, \ v_{\ell})(v_{\ell}-v),
\end{eqnarray}
\begin{eqnarray}
\label{a_syn}
\tilde a^{\rm (syn)}(g, \ v, \ v_{\ell})=A\min(V^{\rm (syn)}_{\rm max}(g)-v, \ 0)+ \nonumber  \\
K(v, \ v_{\ell})(v_{\ell}-v), 
\end{eqnarray}
\begin{equation}
\label{a_jam}
\tilde a^{\rm (jam)}(v)= -K^{\rm (jam)}v.
\end{equation}
Here  
 $V^{\rm (syn)}_{\rm max}(g)$ 
is a gap-dependent maximum vehicle speed in   synchronized flow;
 $K^{\rm (jam)}$ is a sensitivity.

\subsection{Safety Conditions}

 Safety deceleration with a deceleration $a_{\rm s}$ can be applied, if  the vehicle speed  becomes higher than
 the  safe speed $v_{\rm s}(g, \ v_{\ell})$. We use    safety deceleration found from the condition:
\begin{eqnarray}
\label{a_safety}
a_{\rm s}(g, \ v, \ v_{\ell})=A_{\rm s}(v_{\rm s}(g, \ v_{\ell})-v),
\end{eqnarray}
where  $A_{\rm s}$ is the
sensitivity related to safety requirements.
 
 The speed $v_{\rm s}(g, \ v_{\ell})$ in (\ref{a_safety}) is
found
based on the safety condition of Gipps~\cite{Gipps}:
\begin{eqnarray}
\label{Gipps_safety}
v_{\rm s} T_{\rm s}+ v^{2}_{\rm s}/(2b_{\rm s}) \leq
g+v^{2}_{\ell}/(2b_{\rm s}),
\end{eqnarray}
where $T_{\rm s}$ is a safety time gap, $b_{\rm s}$ is a constant
deceleration.\footnote{Note that Eqs.~(\ref{acc}) of the
ATD-model can also be written without the term $a^{\rm (jam)}$ as follows
\begin{eqnarray}
\label{acc_new}
\frac{da}{dt}=\left\{
\begin{array}{ll}
(a^{\rm (free)}- a)/\tau &  \textrm{at $g > G$},  \\
(a^{\rm (syn)}-a)/\tau &  \textrm{at $g \leq G$}.
\end{array} \right.
\end{eqnarray}
In (\ref{acc_new}),  the speed $v_{\rm s}(g, \ v_{\ell})$ in (\ref{a_safety})  is found
based on
the Gipps-condition
(\ref{Gipps_safety}) when $g \geq g^{\rm (jam)}_{\rm max}$ and the speed
$v_{\rm s}(g, \ v_{\ell})=0$ when $g < g^{\rm (jam)}_{\rm max}$.
In the latter case, the formula (\ref{a_safety}) with $v_{\rm s}(g, \
v_{\ell})=0$
plays the role of vehicle deceleration within the wide moving jam phase.}
We use an approximated formula for $v_{\rm s}(g, \ v_{\ell})$
  derived from (\ref{Gipps_safety}) in \ref{A}, which
  enables us to write
 $a_{\rm s}(g, \ v, \ v_{\ell})$ (\ref{a_safety}) as follows
\begin{eqnarray}
\label{a_safety_app}
a_{\rm s}(g, \ v, \ v_{\ell})=
A^{\rm (g)}_{\rm s}(v_{\ell})(g/T_{\rm s}-v)+  K_{\rm
s}(v_{\ell})(v_{\ell}-v),
\end{eqnarray}
where 
\begin{eqnarray}
\label{a_safety_coef}
A^{\rm (g)}_{\rm s}(v_{\ell})=A_{\rm s}T_{\rm s}(T_{\rm s}+v_{\ell}/(2b_{\rm
s}))^{-1}, \\
K_{\rm s}(v_{\ell})=A_{\rm s}(T_{0}+ v_{\ell}/(2b_{\rm s}))(T_{\rm
s}+v_{\ell}/(2b_{\rm s}))^{-1},
\label{a_safety_coef1}
\end{eqnarray}
$T_{0}$ is a constant.

\subsection{Physics of Driver Time Delays \label{time_del_sect}}

In Eqs. (\ref{acc}), the time delay $\tau$ is chosen as 
\begin{eqnarray}
\label{tau}
\tau =\left\{
\begin{array}{ll}
\tau_{\rm s} &  \textrm{at $a_{\rm s} < \min( 0, \ \max (\tilde a^{\rm (phase)}, \ a_{\rm min}), \ a)$}, \\
\tilde \tau &  \textrm{otherwise}.
\end{array} \right. 
\end{eqnarray}
Here, $\tau_{\rm s}$ is a short driver time delay associated with a finite driver reaction time that must be taken into account in the cases when
the driver should decelerate unexpectedly to avoid collisions; 
 $\tilde \tau$  is a time delay in other traffic situations, which
 is chosen different depending on whether
the vehicle accelerates or decelerates:
\begin{eqnarray}
\label{tau_expect}
\tilde \tau =\left\{
\begin{array}{ll}
\tau^{\rm (acc)} &  \textrm{at $a > 0$}, \\
\tau^{\rm (dec)} &  \textrm{at $a \leq 0$}.
\end{array} \right. 
\end{eqnarray}
In turn,  $\tau^{\rm (acc)}$ and $\tau^{\rm (dec)}$ in (\ref{tau_expect}) depend on
the acceleration $a$:
\begin{eqnarray}
\label{tau_a}
\tau^{\rm (acc)} =\left\{
\begin{array}{ll}
\tau^{\rm (acc)}_{0} &  \textrm{at $a <   a^{\rm (phase)}$}, \\
\tau^{\rm (acc)}_{1} &  \textrm{ otherwise},
\end{array} \right. 
\end{eqnarray}
\begin{eqnarray}
\label{tau_d}
\tau^{\rm (dec)} =\left\{
\begin{array}{ll}
\tau^{\rm (dec)}_{0} &  \textrm{at $a \geq   a^{\rm (phase)}$}, \\
\tau^{\rm (dec)}_{1} &  \textrm{otherwise}.
\end{array} \right. 
\end{eqnarray}

The driver time delays $\tau^{\rm (acc)}_{0}$, $\tau^{\rm (dec)}_{0}$, $\tau^{\rm (dec)}_{1}$, 
and $\tau^{\rm (acc)}_{1}$ in (\ref{tau_a}), (\ref{tau_d}) are associated with human expectation of local driving conditions,
in particular, with
spatial and temporal anticipation of a driver in accordance with local adaptation to those traffic situations
in which the driver takes into account both the current and expected future
behaviour of  many vehicles ahead (see also Sect.~\ref{D_B_A}).

$\tau^{\rm (acc)}_{0}$ is the mean time delay when a driver
starts to accelerate or wants to increase the acceleration. This can often occur at the downstream front of a
wide moving jam or a synchronized flow region, i.e., when 
the speed in traffic flow 
downstream of the vehicle is higher than the current vehicle speed. In these cases,  after the preceding vehicle has
begun to accelerate,
the driver also begins to accelerate, however, after a time delay to have a desired time gap to the preceding vehicle.

$\tau^{\rm (dec)}_{0}$ is the mean time delay when the driver
starts to decelerate or wants to decelerate harder in cases in which the driver approaches a region of a lower speed
downstream.

$\tau^{\rm (acc)}_{1}$   corresponds to situations in which the driver
accelerates currently but wants either to stop the acceleration or to reduce it. Thus, $\tau^{\rm (acc)}_{1}$ is the mean driver time delay
in  interruption or reduction of driver acceleration in  cases in which
 the driver recognizes that current  acceleration is greater
than a desired acceleration in the current driving situation.

$\tau^{\rm (dec)}_{1}$   corresponds to situations in which the driver decelerates currently but wants either
to stop the deceleration  or to reduce it. Thus, $\tau^{\rm (dec)}_{1}$  is the mean time delay
in  interruption or reduction of driver deceleration in  cases in which
 the driver recognizes that  current  deceleration is more negative
than a desired deceleration in the current driving situation. 

 \subsection{Model of  Road with On-Ramp Bottleneck \label{On_ramp_M} }

 $\lq\lq$Open" boundary conditions are applied on the main road of the length $L_{0}$.
At the beginning of the road 
free flow conditions  are generated for each vehicle one after another 
at equal time intervals $\tau_{\rm in}=1/q_{\rm in}$ where  $q_{\rm in}$ is the flow rate in the 
incoming boundary flow.
To satisfy safety conditions, a new vehicle appears
only if the distance from the beginning of the road ($x=x_{\rm b}$)
to the position $x=x_{\ell}$ of the farthest upstream vehicle in the lane
 exceeds the distance $v_{\ell} T_{\rm s}+d$. 
The speed  $v$ and coordinate $x$ of a new vehicle  are $v= v_{\ell}$ and $x=x_{\rm b}$,
respectively.
After a vehicle has reached the end of the road, 
it is removed;
before this, the farthest downstream vehicle maintains its speed. 
In the initial state ($t=0$), all vehicles have the same initial speed $v=V^{\rm (free)}(g)$ and space gap $g$,
and $q_{\rm in}=v/(g+d)$.

An on-ramp bottleneck   on the main road
is considered. 
The on-ramp consists of two parts: 
 (i) The merging region of the length  $L_{\rm m}$ that begins at $x=x_{\rm on}$.
 Within this region, vehicles can  
merge onto the main road from the on-ramp.
 (ii)  The part of the on-ramp lane of length $L_{\rm r}$ upstream of the merging region
at which vehicles
move according to  the model equations for a homogeneous  road 
with the maximum speed  $v_{\rm free, \ on}= 90$ km/h. 
At the beginning 
of the 
on-ramp lane  the flow rate to the on-ramp $q_{\rm on}$ 
is given as the flow rate on the main road $q_{\rm in}$.

The following rules are applied for vehicle merging within the merging region.
A speed $\hat v$ is calculated corresponding to formula
\begin{equation} 
\hat v= \min( v^{ +},  \ v+\Delta v^{(1)}_{\rm r})
\label{new_speed_squeezing}
\end{equation}
and then it is used in the merging rules
\begin{equation}
g^{+} >g^{\rm (on)}_{\rm min}+\gamma \hat v T_{\rm s},
 \quad g^{-} >g^{\rm (on)}_{\rm min}+\gamma v^{-} T_{\rm s}.
\label{merging_rules}
\end{equation}
Here
superscripts  $ g^{+}$  and $ g^{-}$ are space gaps
to the preceding vehicle and the trailing vehicle
on the main road, respectively; $ v^{+}$ and $ v^{-}$  are  speeds of
the preceding vehicle and the trailing vehicle, respectively;
$\gamma$, $g^{\rm (on)}_{\rm min}$ and $\Delta v^{(1)}_{\rm r}$ are constants, where
$g^{\rm (on)}_{\rm min}$ is the minimum gap at which vehicle merging is possible,
$\Delta v^{(1)}_{\rm r}$ describes the maximum possible increase
in  speed after vehicle merging.
Note that the finite increase $\Delta v^{(1)}_{\rm r}$ 
 in the vehicle speed (\ref{new_speed_squeezing}) is used to simulate
a  complex driver behaviour during merging onto the main road, especially 
in  synchronized flow: In some cases, before  merging
the driver has to accelerate abruptly,  
to adjust the speed to the speed of the preceding vehicle.

If the conditions  (\ref{merging_rules}) are
satisfied, then    
the vehicle merges onto the main road. 
After merging the vehicle speed $v$ is set to $\hat v $
 (\ref{new_speed_squeezing}) and the vehicle coordinate does not change.
If the conditions (\ref{merging_rules}) are not satisfied,
the vehicle does not merge onto the main road. In this case, 
the vehicle moves in the on-ramp lane until it comes to a stop at
 the end of the merging region.

\subsection{Model Functions and Parameters
\label{parameters}}

Model functions and parameters are shown in Tables~\ref{tableATD1} and~\ref{tableATD2}, respectively.
As explained in Sect.~\ref{D_B_A},
  driver time delays $\tau^{\rm (dec)}_{1}$ 
and $\tau^{\rm (acc)}_{1}$ are chosen to be functions of the vehicle speed; additionally, 
   the synchronization gap $G(v, \ v_{\ell})$ and   sensitivity   $K(v, \ v_{\ell})$
  are chosen to be asymmetric
speed functions depending on whether the vehicle speed $v$ is higher or lower than   the speed  
$v_{\ell}$. Explanations of the function $K(v, \ v_{\ell})$ have been made in item (vii) and (viii) of Sect.~\ref{D_B_A}.

Speed dependence and an asymmetric function for the synchronization gap $G(v, \ v_{\ell})$ are explained by driver behaviour
 as follows. The synchronization gap
 is the   space gap  at which a driver  adapts its speed to the speed of the preceding vehicle.
 Firstly, the synchronization gap is an increasing function of  speed: The lower the speed, the smaller the
 maximum gap   at which the driver can comfortably move in synchronized flow. 
 Secondly, if $v<v_{\ell}$, the driver accelerates and he/she  can start
    speed adaptation at a smaller space gap  than in the opposite case
 $v>v_{\ell}$.
 The function $T^{\rm (syn)}(v)$ is used to have a difference in vehicle space gap in steady states of free flow and synchronized flow
 at a given flow rate.
 This space gap difference, which is used for simulation of a
 first-order F$\rightarrow$S transition, tends towards zero when the density in free flow  approaches 
 the maximum point for free flow $\rho^{\rm (free)}_{\rm max}$ 
 (figures~\ref{Steady_states} (a) and (c)); see also item (v) of 
 Sect.~\ref{D_B_A}).

\begin{table}
\centering
\caption{ATD-model functions}
\label{tableATD1}
\begin{tabular}{l}
\hline\noalign{\smallskip}
\multicolumn{1}{c}{Synchronization gap}\\
\noalign{\smallskip}\hline\noalign{\smallskip}
$G(v, \ v_{\ell})=v \max\big(0, \ T^{\rm (syn)}(v)+ \kappa(v, \ v_{\ell})(v-v_{\ell}) \big)$ \\
$\kappa(v, v_{\ell})=\left\{
\begin{array}{ll}
\kappa^{\rm (acc)} &  \textrm{at $v<v_{\ell}$} \\
\kappa^{\rm (dec)} &  \textrm{at $v\geq v_{\ell}$} \\
\end{array} \right.$ \\
$T^{\rm (syn)}(v)=T^{\rm (syn)}_{0}(1-0.85(v/V_{0})^{2})$ \\
\hline\noalign{\smallskip}
\multicolumn{1}{c}{Sensitivities}\\
\noalign{\smallskip}\hline\noalign{\smallskip}
$K(v, \ v_{\ell})=\left\{
\begin{array}{ll}
K^{\rm (acc)} &  \textrm{at $v<v_{\ell}$} \\
K^{\rm (dec)} &  \textrm{at $v\geq v_{\ell}$}
\end{array} \right.$ \\ 
$K^{\rm (dec)}(v)=K^{\rm (dec)}_{1}(1-\lambda(v))+K^{\rm (dec)}_{2}\lambda(v)$ \\
$\lambda(v)=(1+\exp((v/v_{\rm c}-1)/\epsilon))^{-1}$ \\
\noalign{\smallskip}\hline\noalign{\smallskip}
\multicolumn{1}{c}{Characteristic speed functions}\\
\noalign{\smallskip}\hline\noalign{\smallskip}
$V^{\rm (free)}(g)=V(g)$ \\
$V^{\rm (syn)}_{\rm max}(g)=V(g) $ \\
$V(g)=V_{0} \tanh ((g+2)/(V_{0}T))$ \\
\noalign{\smallskip}\hline\noalign{\smallskip}
\multicolumn{1}{c}{Speed dependensies of time delays}\\
\noalign{\smallskip}\hline\noalign{\smallskip}
$\tau^{\rm (dec)}_{1}(v)=\left\{
\begin{array}{ll}
0.5 \ {\rm s} &  \textrm{at $v \geq v_{\rm c} $} \\
0.7 \ {\rm s} &  \textrm{otherwise}
\end{array} \right.$ \\
$\tau^{\rm (acc)}_{1}(v)=\left\{
\begin{array}{ll}
0.57 \ {\rm s} &  \textrm{at $v \geq v_{\rm c} $} \\
0.87 \ {\rm s} &  \textrm{otherwise}
\end{array} \right.$ \\ 
\hline\noalign{\smallskip}
\end{tabular}
\end{table}

\begin{table}
\centering
\caption{ATD-model parameters}
\label{tableATD2}
\begin{tabular}{l}
\hline\noalign{\smallskip}
$V_{\rm 0}=$ 33.3 m/s (120 km/h), $T=$ 0.9 s, \\
$A= \ 0.5 \ {\rm s^{-1}}$,
$K^{\rm (acc)}= \ 0.8 \ {\rm s^{-1}}$,  $K^{\rm (jam)}=1 \ {\rm s^{-1}}$, \\
$K^{\rm (dec)}_{1}= \ 0.95 \ {\rm s^{-1}}$, $K^{\rm (dec)}_{2}= \ 0.48 \ {\rm s^{-1}}$, 
$v_{\rm c}=$ 15 m/s, $\epsilon=$ 0.15,  \\
 $T^{\rm (syn)}_{0}=$ 2.5 s, 
$\kappa^{\rm (acc)}= \ 0.5 \ {\rm s^{2}/m}$, $\kappa^{\rm (dec)}= \ 0.55 \ {\rm s^{2}/m}$, \\
$g^{\rm (jam)}_{\rm max} = $ 0.95 m, 
$\tau^{\rm (dec)}_{0}=$ 1 s, $\tau^{\rm (acc)}_{0}=$ 0.75 s, 
$\tau_{\rm s}=$ 0.4 s, \\
$a_{\rm max}= \ 1 \ \rm m/s^{2}$, $a_{\rm min}= \ -1 \ {\rm m/s^{2}}$, \\
$A_{\rm s}= \ 1.25 \ \rm s^{-1}$,
$b_{\rm s}=\ 2 \ \rm m/s^{2}$, 
$T_{\rm s}=$ 1 s, $T_{0}=$ 0.42 s, \\
$L_{0}=$ 25 km, $x_{\rm b}= -$ 5 km, $x_{\rm on}=$ 16 km, \\
 $L_{\rm m}=$ 300 m,  
$L_{\rm r}=$ 500 m,   $\gamma=$ 0.22,  $\Delta v^{(1)}_{\rm r}=$ 8 m/s, $g^{\rm (on)}_{\rm min}=$ 0. \\
\hline\noalign{\smallskip}
\end{tabular}
\end{table}

 \subsection{Steady States
\label{SteadyStates}}

In steady states, 
all vehicles have the same speed $v = v_{\ell}$ and the same space gap $g$, and
all accelerations and their time derivatives are zero, and
 the density $\rho$ and the flow rate $q$ are related to the space gap $g$
and to the speed $v$ by the obvious conditions  
\begin{equation}
\rho = 1/(x_{\ell} - x)= 1/(g + d), \quad q = \rho v = v/(g+d).
\label{rho}
\end{equation} 
 According to (\ref{speed})--(\ref{a_safety}) and formulae for $V^{\rm (free)}(g)$, $V^{\rm (syn)}_{\rm max}(g)$
(Table~\ref{tableATD1}) for steady states, we get
\begin{equation}
v  = V(g)    \quad {\rm at} \   g> G(v)
 \ {\rm and} \ g > g^{\rm (jam)}_{\rm max}, 
\label{steady_free}
\end{equation}
\begin{equation}
v  \leq V(g)    \quad {\rm at} \  g \leq G(v)
\ {\rm and} \ g > g^{\rm (jam)}_{\rm max}, 
\label{steady_syn}
\end{equation}
\begin{equation}
v=0    \quad {\rm at} \quad g \leq g^{\rm (jam)}_{\rm max}, 
\label{steady_jam}
\end{equation}
 \begin{equation}
v\leq v_{\rm s}(g, \ v).
 \label{Safe_st}
 \end{equation}

\begin{figure*}
\begin{center}
\includegraphics[width=12 cm]{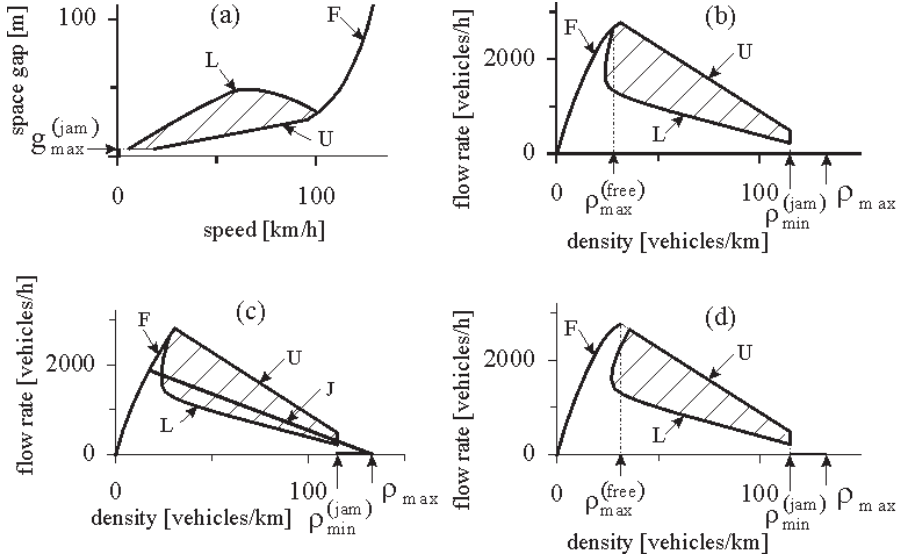}
\caption{Model steady states for the  ATD-models: (a) - In the space-gap--speed plane.
(b) - In the flow--density plane. (c) -- Steady states and the line $J$
(explanation of the line $J$ see in the book~\cite{KernerBook}). (d) -- ATD-model
with  separated steady states in free flow and synchronized flow. 
\label{Steady_states} } 
\end{center}
\end{figure*}

According to (\ref{steady_free})--(\ref{Safe_st}), the model steady states 
consist of the curve $v  = V(g)$ (\ref{steady_free})
  at $g \geq g^{\rm (free)}_{\rm min}$ (curve $F$ in figure~\ref{Steady_states} (a)) 
  for   free flow, a two-dimensional region in the space-gap--speed plane 
 for   synchronized flow  determined by
inequalities in (\ref{steady_syn}), (\ref{Safe_st}), and the line 
$v  = 0$ at $g \leq g^{\rm (jam)}_{\rm max}$ (\ref{steady_jam}) for   wide moving jams (figure~\ref{Steady_states} (a)).
$g^{\rm (free)}_{\rm min}$ is the minimum space gap in free flow
found  as a solution of the set of the equations 
$v  = V(g) \ {\rm and} \  g = G(v,v_{\ell})$ at $v=v_{\ell}$.

The two-dimensional region for steady states of synchronized flow 
is limited by the following boundaries: 
the  boundary $U$, the  curve $L$, the curve $v  = V(g)$ 
  at $g < g^{\rm (free)}_{\rm min}$, and 
the horizontal line $g = g^{\rm (jam)}_{\rm max}$.
 The   boundary $U$ is associated with the safe speed, i.e., this is determined by the  condition (\ref{Safe_st}) when it is
 an equality.
This leads to the condition for the boundary $U$
 \begin{equation}
g = vT_{\rm s}.
 \label{Safe_st2}
 \end{equation}
The   boundary $L$  is found from the condition that
the vehicle space  gap is equal to the synchronization gap
 \begin{equation}
g=G(v).
 \label{gap_st}
 \end{equation}

In the flow--density plane, free flow (curve $F$ in figure~\ref{Steady_states} (b))
is found from 
\begin{equation}
q= \rho V_{\rm F}(\rho) 
\label{left}
\end{equation}
at $\rho \leq \rho^{\rm (free)}_{\rm max}$
where $V_{\rm F}(\rho)=V(g)\mid_{g=\rho^{-1}-d}$,
 $\rho^{\rm (free)}_{\rm max}=(g^{\rm (free)}_{\rm min}+d)^{-1}$.
 A wide moving jam is associated with the horizontal line
$q= 0$  at $\rho^{\rm (jam)}_{\rm min} \leq \rho \leq \rho_{\rm max}$
(figure~\ref{Steady_states} (b)),
 where $\rho^{\rm (jam)}_{\rm min}=(g^{\rm (jam)}_{\rm max}+d)^{-1}$,
 $\rho_{\rm max}=d^{-1}$. The boundaries of a two-dimensional region for steady states of
 synchronized flow are:
the upper line $U$ determined by the condition $q =  (1-\rho d)/T_{\rm s}$, the lower curve $L$
determined by the condition $\rho G(q/\rho) = 1-\rho d$,  the curve (\ref{left})
  at $\rho > \rho^{\rm (free)}_{\rm max}$, and 
the vertical line $\rho = \rho^{\rm (jam)}_{\rm min}$.\footnote{We have also studied another version of the ATD-model
 in which there is a separation of steady states in free flow and synchronized flow
 in the flow--density plane, i.e., the maximum point for free flow $\rho^{\rm (free)}_{\rm max}$
 is related to the intersection point of the line $U$
 and the curve $F$ (figure~\ref{Steady_states} (d)). Simulations of this version of the ATD-model
 show qualitatively the same features of phase transitions and congested patterns as those discussed in
 Sect.~\ref{PT_Onramp}.}

\section{Speed Adaptation Model}
\label{SA_Section}

\subsection{Empirical F$\rightarrow$S$\rightarrow$J Transitions as Physical Basis of Speed Adaptation Model}

\begin{figure*}
\begin{center}
\includegraphics[width=12 cm]{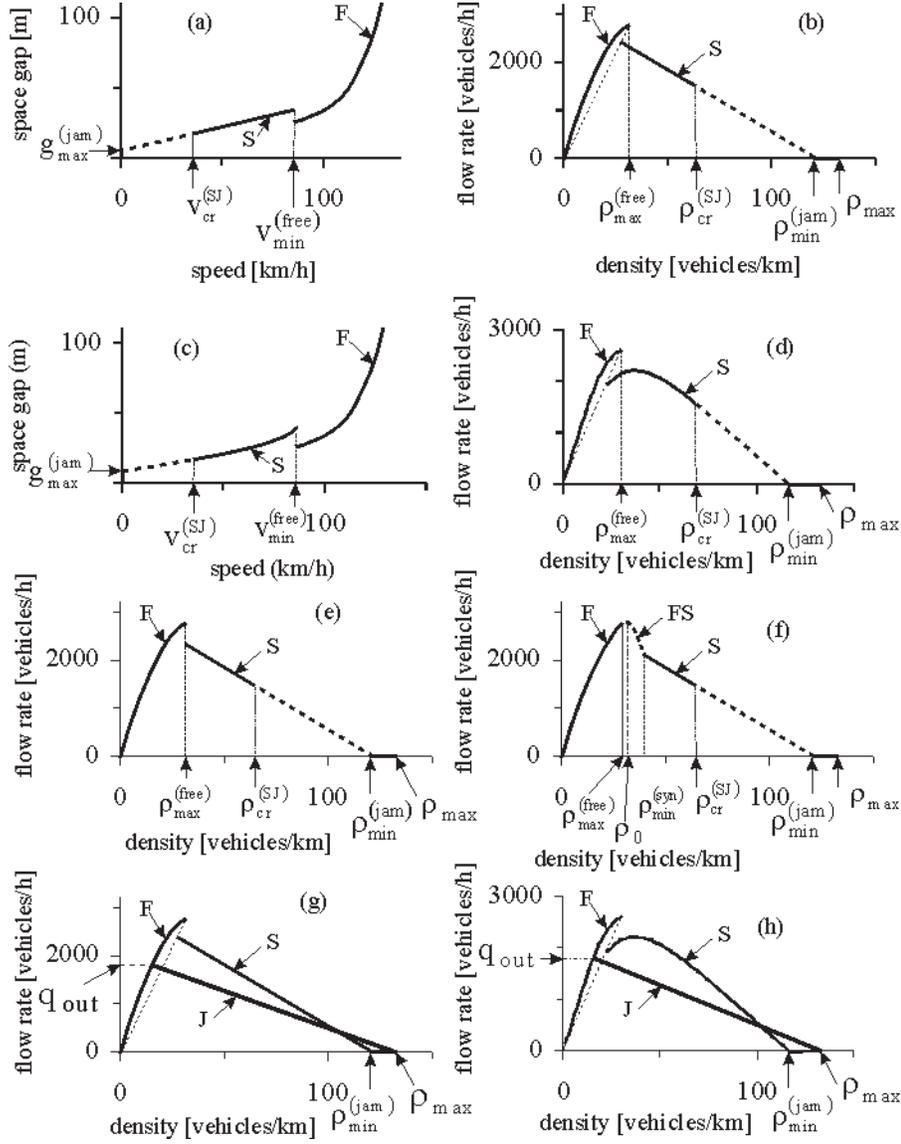}
\caption{Model steady states for   SA-models: (a, b) - In the space-gap--speed (a)
and  the flow--density plane  (b) for the  SA-model (\ref{coor_sa})--(\ref{v_syn_sa_2}). (c, d) - The   
SA-model (\ref{coor_sa})--(\ref{a_jam_sa}), (\ref{v_syn_sa_1}).
(e, f) - Other possible SA-models (see~\ref{B}).   Dashed parts of curves for model steady states in (a--f)
are associated with unstable model steady states. (g, h) -- Steady states and the line $J$
for the SA-model (a, b) and the SA-model (c, d).
\label{Steady_states_SA} } 
\end{center}
\end{figure*}

 The fundamental hypothesis of three-phase traffic theory, which postulates that
hypothetical steady states of synchronized flow cover a two-dimensional region in the flow--density plane, is also 
one of the basic hypotheses of the ATD-model presented above (figures~\ref{Steady_states} (a) and (b)). In contrast with the ATD-model,
in a speed adaptation model   ({\bf s}peed {\bf a}daptation model, SA-model for short) 
hypothetical steady states of synchronized flow are associated with a curve (curve $S$ in figures~\ref{Steady_states_SA} (a) and (b)), i.e., they
cover a one-dimensional region in the flow--density plane. The curve $S$
 is associated with an averaging of an infinite number of steady states of synchronized flow to one synchronized flow speed 
 for each vehicle space gap. A gap dependence of the average speed in synchronized flow steady   states on the curve $S$ is
  denoted by $V^{\rm (syn)}_{\rm av}(g)$. The basis hypothesis of the SA-model is associated with
 the sequence of F$\rightarrow$S$\rightarrow$J transitions,
 which determine moving jam emergence in empirical observations~\cite{Kerner1998B,KernerBook}.
 
 Note that as in the models and theories in the context
 of the fundamental diagram approach~\cite{Gartner,Sch,Helbing2001,Nagatani_R,Nagel2003A}, in the
 SA-model steady state model solutions cover a one-dimension region(s)
 in the flow--density plane. However, in    the  models and theories 
 reviewed~\cite{Gartner,Sch,Helbing2001,Nagatani_R,Nagel2003A},
 which claim to show spontaneous moving jam emergence, 
    the F$\rightarrow$J transition governs the onset of congestion. 
    This is inconsequent with empirical results~\cite{KernerBook}. 
 In contrast, 
   in the SA-model the onset of congestion  is associated with an
 F$\rightarrow$S transition, whereas moving jams  occur spontaneously only  in synchronized flow,
 in accordance with empirical results. 
 
 The SA-model
  is   simpler than the ATD-model.
However, due to this simplification the SA-model
cannot show some features of congested patterns  of the ATD-model
(Sect.~\ref{SA_Con_Sec}), which are observed in empirical observations.
 The purpose of the
 SA-model is  to simulate an F$\rightarrow$S transition and    features of
  the sequence of 
  F$\rightarrow$S$\rightarrow$J transitions,
as observed in empirical observations~\cite{Kerner1998B,KernerBook}, in a   simple way.  
This confirms an assumption of
 three-phase traffic theory  that if rather than the fundamental hypothesis the hypothesis about the 
  F$\rightarrow$S$\rightarrow$J transitions
is the basis of a mathematical model, then the model can show and predict some important
empirical features of the phase transitions (see footnote 4 of Sect. 4.3.4 in~\cite{KernerBook}).
 
In the SA-models, an F$\rightarrow$S transition is modelled
through two effects: (i) Discontinuouty of steady speed solutions (figures~\ref{Steady_states_SA} (a), (c), and (e)) 
or their instability (curve $FS$ in figure~\ref{Steady_states_SA} (f))
in the vicinity of the maximum point of free flow $v^{\rm (free)}_{\rm min}$, $\rho^{\rm (free)}_{\rm max}$.
(ii) The speed adaptation effect is modelled through the term   
$K(v, \ v_{\ell})(v-v_{\ell})$  that  adjusts the speed to the preceding vehicle in synchronized flow.

Moving jam emergence is simulated  through an instability of  some of the synchronized flow model steady states
associated with the curve $V^{\rm (syn)}_{\rm av}(g)$.
This instability occurs in synchronized flow    at lower   speeds
 and greater densities (i.e., smaller space gaps).
The associated   critical density and speed of the synchronized flow steady states are denoted by $\rho^{\rm (SJ)}_{\rm cr}$
and $v^{\rm (SJ)}_{\rm cr}$, respectively
(figure~\ref{Steady_states_SA}). 
To simulate this instability, as in the ATD-model (item (viii) of Sect.~\ref{D_B_A}), in the SA-models
the sensitivity $K(v, \ v_{\ell})$ at $v\geq v_{\ell}$ is a decreasing speed function. 
Similarly with the ATD-model, to simulate
the   mean time delay in acceleration at the downstream jam front in the SA-model, 
 a vehicle  within the jam does not accelerate before
 (\ref{g_g_jam})
 is satisfied (item (vii) of Sect.~\ref{D_B_A}).

\subsection{Basic Equations}

There can be   different possibilities for a separation of steady states of free flow  and synchronized flow
in SA-models, which all exhibit qualitatively the same 
features of the F$\rightarrow$S$\rightarrow$J transitions. To illustrate this, here we consider
two variants of  SA-models; in~\ref{B}
  other possible variants of  SA-models are discussed. All these variants of the SA-models
 exhibit very similar features of phase transitions and spatiotemporal 
 congested traffic patterns that are associated with the
 same physics of these SA-models. 

A   formulation for the SA-model
reads as follows
\begin{eqnarray}
\label{coor_sa}
\frac{dx}{dt}=v,  \\
\label{acc_sa}
\frac{dv}{dt}=\left\{
\begin{array}{ll}
a^{\rm (free)} &  \textrm{at $v \geq v^{\rm (free)}_{\rm min}$ and $g > g^{\rm (jam)}_{\rm max}$},  \\
a^{\rm (syn)} &  \textrm{at $v < v^{\rm (free)}_{\rm min}$ and $g > g^{\rm (jam)}_{\rm max}$}, \\
a^{\rm (jam)} &  \textrm{at $0 \leq  g \leq g^{\rm (jam)}_{\rm max}$}.
\end{array} \right. 
\end{eqnarray}
The vehicle acceleration $a=dv/dt$ in (\ref{acc_sa}) is supposed to be limited by
the maximum acceleration $ a_{\rm max}$, i.e., in (\ref{acc_sa}) 
\begin{eqnarray}
a^{(\rm phase)}= \min(\tilde a^{\rm (phase)}, \ a_{\rm max}).
\label{acc_lim_sa}
\end{eqnarray}
Here and below the associated designations of functions and parameters have the same meaning as those in the ATD-model
(Sect.~\ref{TimeDelayS}).

\subsection{Vehicle Acceleration}

Functions 
$\tilde a^{\rm (free)}(g, \ v, \ v_{\ell})$, $\tilde  a^{\rm (syn)}(g, \ v, \ v_{\ell})$,  and $\tilde  a^{\rm (jam)}(v)$ 
in (\ref{acc_lim_sa})
are chosen as follows
\begin{eqnarray}
\label{a_free_sa}
\tilde a^{\rm (free)}(g, \ v, \ v_{\ell})=
A^{\rm (free)}(V^{\rm (free)}(g)-v)+  \nonumber \\
K(v, \ v_{\ell})(v_{\ell}-v),
\end{eqnarray}
\begin{eqnarray}
\label{a_syn_sa}
\tilde a^{\rm (syn)}(g, \ v, \ v_{\ell})=A^{\rm (syn)}
\big(V^{\rm (syn)}_{\rm av}(g)-v\big)+ \nonumber  \\
K(v, \ v_{\ell})(v_{\ell}-v), 
\end{eqnarray}
\begin{equation}
\label{a_jam_sa}
\tilde a^{\rm (jam)}(v)= -K^{\rm (jam)} v.
\end{equation}
 
Two versions  of functions $V^{\rm (syn)}_{\rm av}(g)$ in (\ref{a_syn_sa}) that lead
to two different versions of the SA-models are considered:
\begin{equation}
\label{v_syn_sa_2}
V^{\rm (syn)}_{\rm av}(g)=\tilde g(g)/T^{\rm (syn)}_{\rm av},
\end{equation}
and
\begin{equation}
\label{v_syn_sa_1}
V^{\rm (syn)}_{\rm av}(g)=V_{1}\bigg[\tanh\bigg(\frac{\tilde g(g)}{T^{\rm (syn)}_{\rm av}V_{1}}\bigg)+c\tilde g(g)\bigg],
\end{equation}
where $\tilde g(g)= g-g^{\rm (jam)}_{\rm max}$; 
$T^{\rm (syn)}_{\rm av}$, $V_{1}$ and $c$ are constants.

\subsection{Steady States and Model Parameters \label{Steady_sect_SA}}

In the SA-models, in accordance with (\ref{acc_sa})
there are three isolated curves for steady states
of the SA-models associated with the three traffic phases: free flow, synchronized flow, and wide moving jam
(figures~\ref{Steady_states_SA} (a) and (b)).

Steady states of free flow 
  are related to a curve $v=V_{\rm F}(\rho)$ and formula (\ref{left}) (the curve $F$ 
in figures~\ref{Steady_states_SA} (a)--(d)) associated with the condition
\begin{equation}
v = V^{\rm (free)}(g)    \quad {\rm at} \   v \geq v^{\rm (free)}_{\rm min}.
\label{steady_free_sa}
\end{equation}

Steady states of synchronized flow 
  are related to 
a curve $S$ in the space-gap--speed plane (figures~\ref{Steady_states_SA} (a) and (c))
given by the condition
\begin{equation}
v=V^{\rm (syn)}_{\rm av}(g)
 \quad {\rm at} \   v < v^{\rm (free)}_{\rm min} \ {\rm and} \ g > g^{\rm (jam)}_{\rm max}.
\label{steady_syn_sa}
\end{equation}
In terms of the flow rate $q$ and  density $\rho$, the formula  (\ref{steady_syn_sa}) reads
\begin{equation}
q= \rho V_{\rm S}(\rho) 
 \quad {\rm at} \  \rho^{\rm (syn)}_{\rm min} < \rho< \rho^{\rm (jam)}_{\rm min},
\label{steady_syn0_sa}
\end{equation}
where $V_{\rm S}(\rho)=V^{\rm (syn)}_{\rm av}(g)\mid_{g=\rho^{-1}-d}$,
 $\rho^{\rm (syn)}_{\rm min}=(g^{\rm (syn)}_{\rm max}+d)^{-1}$,
$g^{\rm (syn)}_{\rm max}$ is found from the equation
$V^{\rm (syn)}_{\rm av}(g^{\rm (syn)}_{\rm max})=v^{\rm (free)}_{\rm min}$.

In the case of the function $V^{\rm (syn)}_{\rm av}(g)$ given by (\ref{v_syn_sa_2}),
the  formula (\ref{steady_syn0_sa}) 
yields the equation for a curve $S$ with a negative slope in the flow--density plane (figure~\ref{Steady_states_SA} (b))
\begin{equation}
q =  (1-\rho/ \rho^{\rm (jam)}_{\rm min})/T^{\rm (syn)}_{\rm av}
 \quad {\rm at} \  \rho^{\rm (syn)}_{\rm min} < \rho< \rho^{\rm (jam)}_{\rm min}.
\label{steady_syn0_sa2}
\end{equation}
When the function $V^{\rm (syn)}_{\rm av}(g)$ is given by formula (\ref{v_syn_sa_1}),
the curve $S$  has a maximum in the flow--density plane (figure~\ref{Steady_states_SA} (d)).

Steady states for a wide moving jam 
 are the same as those in the ATD-model, i.e., they
are given by a  horizontal  line 
\begin{equation}
q =0
 \quad {\rm at} \  \rho^{\rm (jam)}_{\rm min} \leq \rho \leq \rho_{\rm max}
\label{steady_jam0_sa}
\end{equation}
in the flow--density plane (figures~\ref{Steady_states_SA} (b) and (d)).

\begin{table}
\centering
\caption{SA-model parameters}
\label{tableSA2}
\begin{tabular}{l}
\hline\noalign{\smallskip}
   $V^{\rm (free)}(g)= V(g)$, \\
$V(g)=V_{0} \tanh (g/(V_{0}T))$,
$V_{0}=33.3$ m/s (120 km/h), $ T=0.85$ s, \\
$a_{\rm max}=2 \ {\rm ms^{-2}}$;
$A^{\rm (free)}=0.4 \ {\rm s^{-1}}$, \\
$A^{\rm (syn)}=0.1 \ {\rm s^{-1}}$,
$K^{\rm (jam)}=2.2 \ {\rm s^{-1}}$, \\
  $K(v, \ v_{\ell})$ is given in Table~\ref{tableATD1}, \\
$K^{\rm (dec)}(v)$ is given in  Table~\ref{tableATD1} at
$v_{\rm c}=10$ m/s, $\epsilon=0.07$, \\
$g^{\rm (on)}_{\rm min}=g^{\rm (jam)}_ {\rm max}$, $\gamma=0.25$. \\
In the  SA model with function $V^{\rm (syn)}_{\rm av}(g)$ (\ref{v_syn_sa_2}), we use \\
$v^{\rm (free)}_ {\rm min}= 22.22$ m/s (80 km/h),
$K^{\rm (dec)}_{1}=0.95 \ {\rm s^{-1}}$$K^{\rm (dec)}_{2}=0.64 \ {\rm s^{-1}}$, \\
$K^{\rm (acc)}=0.4 \ {\rm s^{-1}}$;
$ T^{\rm (syn)}_{\rm av}=1.2$ s,
$g^{\rm (jam)}_ {\rm max}=0.7$ m,
 $\Delta v^{(1)}_{\rm r}=3.5$ m/s. \\
In the  SA model with   function $V^{\rm (syn)}_{\rm av}(g)$  (\ref{v_syn_sa_1}), we use \\
$v^{\rm (free)}_ {\rm min}= 23.61$ m/s (85 km/h),
$K^{\rm (dec)}_{1}=0.95 \ {\rm s^{-1}}$, $K^{\rm (dec)}_{2}=0.75 \ {\rm s^{-1}}$, \\
$K^{\rm (acc)}(v)=0.3 +0.4\min(1,  v/12)  \ {\rm s^{-1}}$, $V_{1}=20$ m/s,\\
 $c=0.007 \ {\rm m^{-1}}$,
$ T^{\rm (syn)}_{\rm av}=1$ s,
$g^{\rm (jam)}_ {\rm max}=1$ m,
 $\Delta v^{(1)}_{\rm r}=2.5$ m/s. \\
\hline\noalign{\smallskip}
\end{tabular}
\end{table}

Parameters of the SA-models are shown in Table~\ref{tableSA2}.

\section{Diagram of Congested Traffic Patterns at On-Ramp Bottleneck in ATD-Model
\label{PT_Onramp}}

\begin{figure*}
\begin{center}
\includegraphics[width=12 cm]{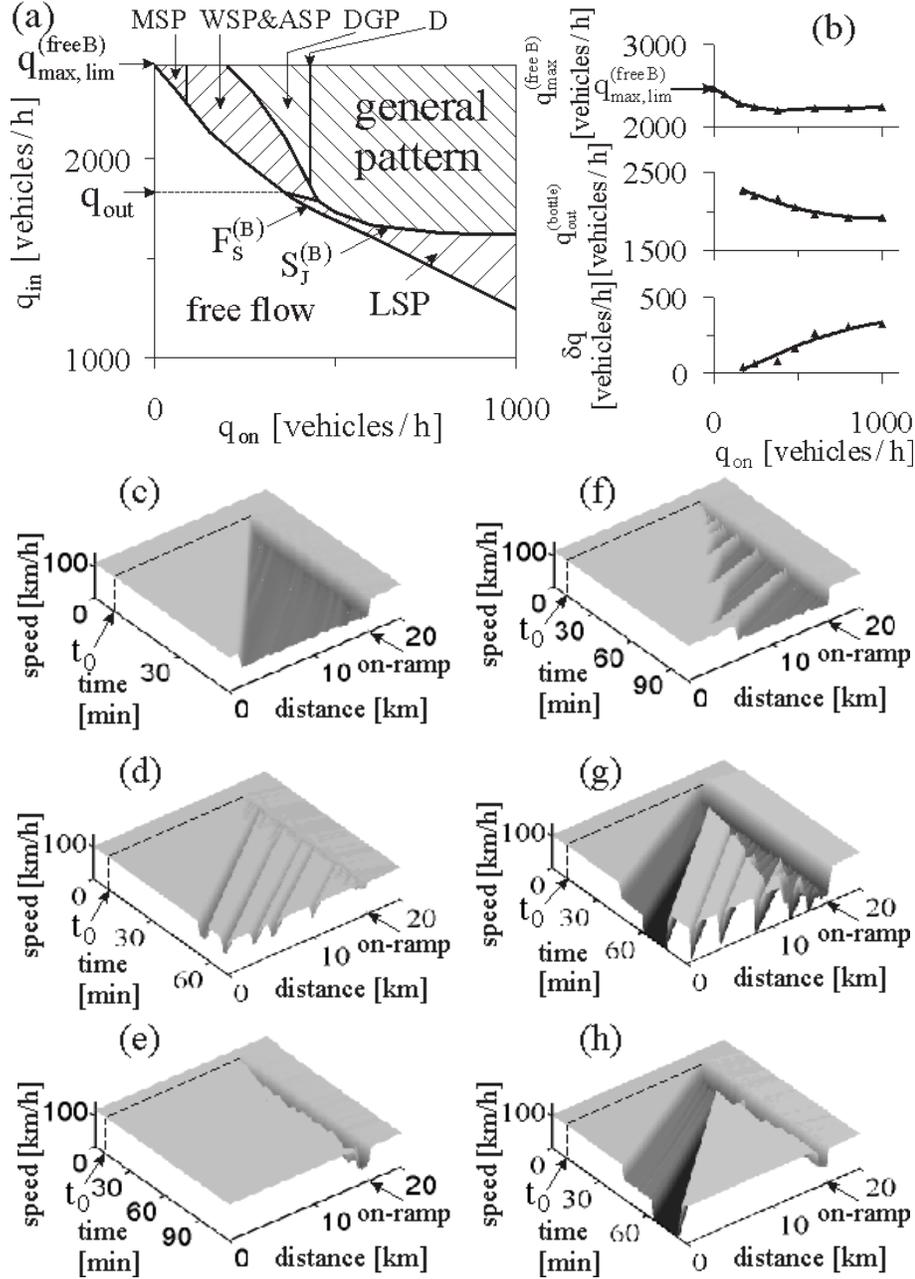}
\caption{Congested patterns in the ATD-model: (a) --
Diagram of congested patterns. (b) -- Maximum capacity in
free flow at the bottleneck $q^{\rm (free \ B)}_{\rm max}$,
 the discharge flow rate downstream of the congested bottleneck $q^{\rm (bottle)}_{\rm out}$,
 and the capacity drop $\delta q$. (c--h) -- 
Congested patterns upstream of the bottleneck related to (a):
(c--f) -- Synchronized flow patterns (SPs) and (g, h) -- general patterns (GPs). 
(c) -- Widening SP (WSP).
(d) -- Sequence of moving SPs (MSPs).
(e) -- Localized SP (LSP).
(f) -- Alternating synchronized flow pattern (ASP).
(g) -- GP.
(h) -- Dissolving GP (DGP).
In (c--h) 
the flow rates 
$(q_{\rm  on}, q_{\rm  in})$ are: 
(c) (350, 2140), 
(d) (60, 2367), 
(e) (360, 1800),
(f) (240, 2026),
(g) (510, 2310), 
(h) (360, 2310) vehicles/h. 
$q^{\rm (free \ B)}_{\rm max, \ lim}\approx 2470 $ vehicles/h.
In (b) the discharge flow rate $q^{\rm (bottle)}_{\rm out}$ is changed
from 2270    to 1925  vehicles/h, 
$q_{\rm out} \approx 1805$ vehicles/h. $T_{\rm ob}=$ 30 min for the boundary  $F^{\rm (B)}_{\rm S}$
and 60 min for the boundary  $S^{\rm (B)}_{\rm J}$. $t_{0}=$ 7 min.
\label{Diagram_ATD} } 
\end{center}
\end{figure*}

Numerical simulations of the
ATD-model show that  congested patterns (figure~\ref{Diagram_ATD}), which appear on the main road upstream of
 the bottleneck,
 are qualitatively the same as those for the stochastic models of Ref.~\cite{KKl,KKW,KKl2003A} reviewed in the book~\cite{KernerBook}.
 However, dynamics of phase transitions leading to  congested pattern formation and a  diagram of these patterns
 in the flow--flow plane with co-ordinates are $q_{\rm on}$ and  $q_{\rm in}$ (figure~\ref{Diagram_ATD} (a))
  exhibit some important peculiarities in comparison with the stochastic models~\cite{KKl,KKW,KKl2003A}. 
  These peculiarities are associated with a deterministic character of the ATD-model.
  To understand this, firstly features of  an F$\rightarrow$S transition at the bottleneck in 
  the deterministic ATD-model should be considered.

\subsection{Local Perturbation and F$\rightarrow$S Transition in Free Flow at Bottleneck  \label{ATD_LP_Sect} }

\begin{figure*}
\begin{center}
\includegraphics[width=12 cm]{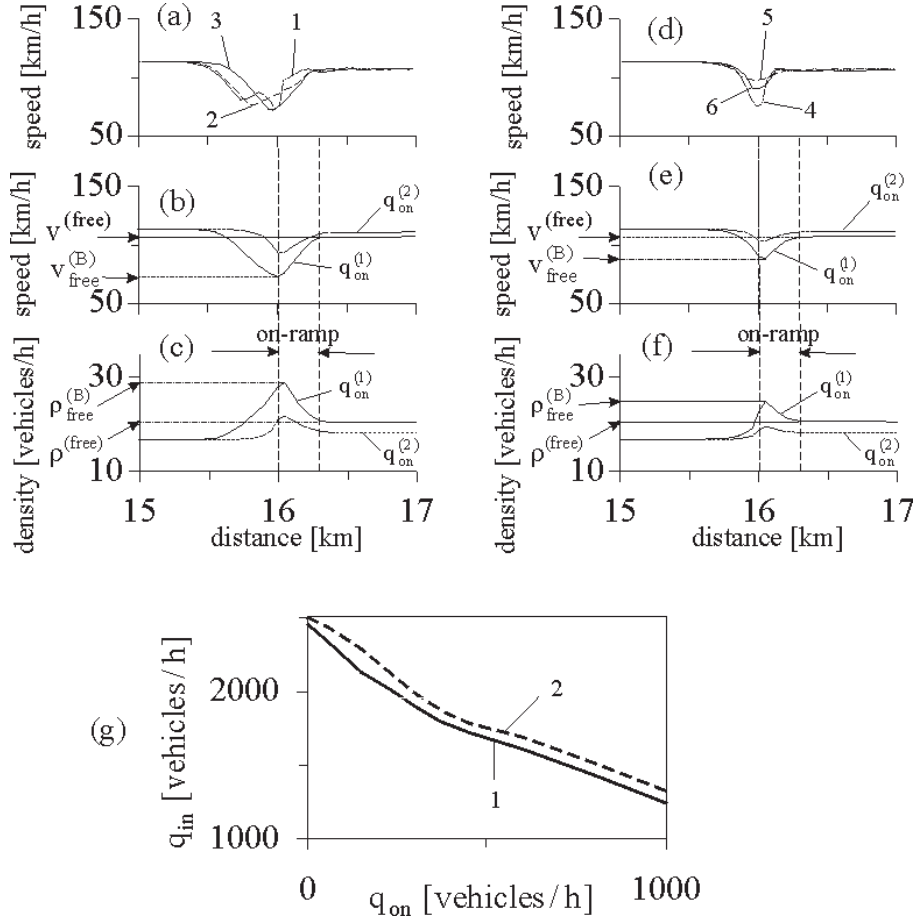}
\caption{Perculiarities of F$\rightarrow$S transitions  in the ATD-model: (a--f) --
Space dependences of speed and density within local perturbations at on-ramp bottleneck at two different parameters of vehicle merging
from on-ramp onto the main road $\Delta v^{(1)}_{\rm r}=$ 8 m/s (a--c) and $\Delta v^{(1)}_{\rm r}=$ 12 m/s
(d--f). 
$q_{\rm on}=q^{(1)}_{\rm on}=$ 300, $q_{\rm on}=q^{(2)}_{\rm on}=$ 120, and $q_{\rm in}=$ 1900 vehicles/h. In (a, d) curves 1--6 are related to 
different time moments.
(g) -- Boundaries $F^{\rm (B)}_{\rm S}$ of a first-order F$\rightarrow$S transition at the bottleneck
associated with (a) and  (d), respectively. The boundary $F^{\rm (B)}_{\rm S}$ 
associated with  curve 1 in (g) is taken from figure~\ref{Diagram_ATD} (a).
In (b, c, e, f), 5-min averaging  of traffic variables
measured at virtual detectors at different locations are shown.
Other model parameters are the same as those in figure~\ref{Diagram_ATD}.
\label{ATD_Pert} } 
\end{center}
\end{figure*}

   Vehicle merging results in a abrupt local  space gap reduction 
    on the main road. This can lead to abrupt local vehicle deceleration.
 For this reason, a dynamic decrease in speed (figure~\ref{ATD_Pert} (a)) and the associated increase in density in the on-ramp merging
 region appear. This local disturbance in the speed and density  localized at the bottleneck can be considered
a time-dependent dynamic perturbation in free flow.  The dynamic nature of this perturbation
(there are no random fluctuations in the deterministic ATD-model) is
explained by dynamic rules of vehicle motion and by a spatial non-homogeneity   localized in the on-ramp merging region
within which
on-ramp inflow and  flow on the main road merge.
If the flow rate $q_{\rm in}$ is great enough, then due to dynamic merging rules of    Sect.~\ref{On_ramp_M}
  vehicles can merge onto the main road 
     at     different locations within the merging region. This complex dynamic vehicle merging behaviour
causes the associated complex dynamic  spatiotemporal dependence of the speed and, respectively, density within the dynamic perturbation
 (figure~\ref{ATD_Pert} (a)).

If    the speed and density within the perturbation are averaged 
over time with an averaging time interval that is considerably longer than time intervals between merging of vehicles, then
spatial distributions of the speed and density within the associated average perturbation  (figures~\ref{ATD_Pert} (b) and (c))
can be considered a $\lq\lq$deterministic"
perturbation localized at on-ramp bottleneck. At this time scale the deterministic perturbation 
 is motionless, the total flow rate (across the main road and  on-ramp lane)
within the perturbation does not depend on
spatial co-ordinate. This total flow rate in free flow is $q_{\rm sum}=q_{\rm in}+q_{\rm on}$.
In contrast, the average speed  and density  spatially
vary in free flow at the bottleneck. In particular,
$q_{\rm sum}=v^{\rm (B)}_{\rm free}\rho^{\rm (B)}_{\rm free}=v^{\rm (free)}\rho^{\rm (free)}$, where  
  $v^{\rm (B)}_{\rm free}$ and  $\rho^{\rm (B)}_{\rm free}$ are the minimum speed and maximum density 
  within the deterministic perturbation, respectively; $v^{\rm (free)}$, $\rho^{\rm (free)}$ are
  the speed and density  downstream of the perturbation, respectively
(figures~\ref{ATD_Pert} (b) and (c));
$v^{\rm (B)}_{\rm free}< v^{\rm (free)}$, $\rho^{\rm (B)}_{\rm free}>\rho^{\rm (free)}$.

At a given $q_{\rm in}$, the greater $q_{\rm on}$, the lower the speed $v^{\rm (B)}_{\rm free}$
and the greater the density $\rho^{\rm (B)}_{\rm free}$ within the perturbation, i.e., the greater the amplitude of the
deterministic perturbation (figures~\ref{ATD_Pert} (b) and (c)).
This growth in the perturbation amplitude has a limit associated with an F$\rightarrow$S transition
that occurs spontaneously at the bottleneck when $q_{\rm on}$ gradually increases.
The multitude of the flow rates $q_{\rm in}$ and $q_{\rm on}$, at which the F$\rightarrow$S transition
occurs, determines the boundary $F^{\rm (B)}_{\rm S}$ in the pattern diagram (figure~\ref{Diagram_ATD} (a)).
At the boundary $F^{\rm (B)}_{\rm S}$ a first-order F$\rightarrow$S transition (see Sect.~\ref{ATD_SA_Meta}) occurs spontaneously
during a chosen time interval $T_{\rm ob}$ that is considerably longer than
a  time interval $\tau^{\rm (grow \ B)}_{\rm determ}$ (about  60 s) required for the average speed to decrease from the speed
 within a dynamic perturbation in free flow at the bottleneck to  
a synchronized flow speed  
(see explanations in Sect. 5.3.7 of~\cite{KernerBook}). The necessity of the time interval $T_{\rm ob}$
is associated with a  time delay $T^{\rm (B)}_{\rm FS}$
for an F$\rightarrow$S transition found in the ATD-model: After the time delay $T^{\rm (B)}_{\rm FS}$, a 
 time-dependent (dynamic) perturbation (figure~\ref{ATD_Pert} (a)), which
can cause a  short-time decrease in the speed within the perturbation  markedly lower than $v^{\rm (B)}_{\rm free}$,
can occur. This perturbation occurrence
 leads to the F$\rightarrow$S transition. The boundary $F^{\rm (B)}_{\rm S}$ is determined from the condition 
  $T_{\rm ob}\approx T^{\rm (B)}_{\rm FS}$.

In stochastic models~\cite{KernerBook}, the boundary $F^{\rm (B)}_{\rm S}$
is also determined by the considition that an F$\rightarrow$S transition
occurs at given $q_{\rm in}$ and $q_{\rm on}$ after a  time delay $T^{\rm (B)}_{\rm FS}$ during a chosen time interval $T_{\rm ob}$.
However, in the stochastic models $T^{\rm (B)}_{\rm FS}$ is a {\it random} value: In different realizations
made at the same $q_{\rm in}$ and $q_{\rm on}$ various $T^{\rm (B)}_{\rm FS}$ are found.
This stochastic model nature enables us also to calculate the probability for  F$\rightarrow$S transition
occurrence~\cite{KKW,KernerBook}.

In contrast with the stochastic models~\cite{KernerBook}, in the deterministic ATD-model there are no random fluctuations.
Time-dependent perturbations in free flow localized at the bottleneck (figure~\ref{ATD_Pert} (a)) have
dynamic nature explained above. For this reason, in the ATD-model
$T^{\rm (B)}_{\rm FS}$ is a {\it fixed} value  at given
 $q_{\rm in}$ and $q_{\rm on}$; consequently, the probability for  F$\rightarrow$S transition
occurrence cannot be found.

In addition, numerical simulations of the ATD-model show that
a duration of a dynamic  speed decrease within the perturbation  below
  the speed     $v^{\rm (B)}_{\rm free}$ is considerably shorter (1--3 s) than
  $\tau^{\rm (grow \ B)}_{\rm determ}$. As a result, it is found that at a given $q_{\rm in}$ the time delay
  $T^{\rm (B)}_{\rm FS}$ is a strong decreasing function of $q_{\rm on}$ in a neightborhood of
    the boundary $F^{\rm (B)}_{\rm S}$: Already a small increase in $q_{\rm on}$ behind the boundary $F^{\rm (B)}_{\rm S}$
    leads to a decrease in $T^{\rm (B)}_{\rm FS}$ down to $\tau^{\rm (grow \ B)}_{\rm determ}$.
    Thus, we can suggest  that
    in the ATD-model the boundary $F^{\rm (B)}_{\rm S}$ is very close to the boundary for the deterministic
    F$\rightarrow$S transition (see explanation of the deterministic F$\rightarrow$S transition  in Sect. 5.3.7 of Ref.~\cite{KernerBook}).

    The dynamic character of perturbations at the bottleneck, which is responsible for  the above
    mentioned physics of the boundary $F^{\rm (B)}_{\rm S}$
    for an F$\rightarrow$S transition in 
     the ATD-model, can clear be seen, if smaller disturbances in speed and density occur due to vehicle merging.
    Smaller disturbances can be simulated by 
    an increase in the parameter $\Delta v^{(1)}_{\rm r}$ of vehicle merging (Sect.~\ref{On_ramp_M}). As a result,
      at the same $q_{\rm in}$ and $q_{\rm on}$
    as those in  figures~\ref{ATD_Pert} (a)--(c) both time-dependent (figure~\ref{ATD_Pert} (d))  and deterministic perturbations
     (figures~\ref{ATD_Pert} (e) and (f)) become
    smaller. This leads to a shift of the boundary $F^{\rm (B)}_{\rm S}$
    in the diagram of congested patterns to  greater $q_{\rm on}$ (curve 2 in figure~\ref{ATD_Pert} (g)):

\subsection{Perculiarities of S$\rightarrow$J Transitions and Congested Patterns   \label{ATD_Diagram_Sect} }

In the ATD-model,
moving jam formation in synchronized flow (S$\rightarrow$J transition), which occurs at the boundary $S^{\rm (B)}_{\rm J}$ in the
congested pattern diagram  (figure~\ref{Diagram_ATD} (a)),
exhibits
 also some qualitative different features in comparison with the stochastic models~\cite{KernerBook}.
 
As in the stochastic models~\cite{KernerBook}, in the ATD-model  after a synchronized flow pattern (SP) 
occurs upstream of the bottleneck due to F$\rightarrow$S transition at the bottleneck,
 a further   increase in $q_{\rm on}$ leads to a subsequent decrease in the speed within the SP. This
   can cause an S$\rightarrow$J transition with the following general pattern (GP) formation.
In the stochastic models,
a self-growth of random model fluctuations is mostly responsible for the S$\rightarrow$J transition.
In contrast, in the ATD-model there are no random model fluctuations.

\begin{figure*}
\begin{center}
\includegraphics[width=12 cm]{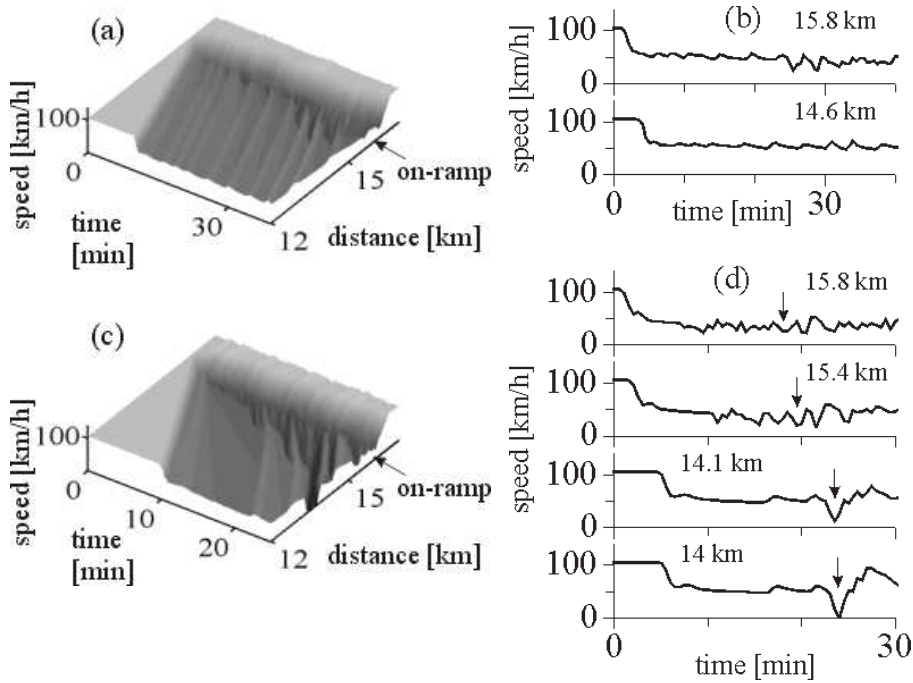}
\caption{Perculiarities of S$\rightarrow$J transitions  in the ATD-model: (a, b) --
Spatiotemporal {\it decay} of  dynamic speed waves that emerge in the merging on-ramp region
during their upstream propagation through a widening SP (WSP) in space and time (a) and at different virtual detectors (b).
(c, d) -- Spatiotemporal {\it growth}
 of  dynamic speed waves that emerge in the merging on-ramp region
during their upstream propagation through an initial WSP in space and time (c) and at different virtual detectors (d). 
$q_{\rm on}=$ 300 (a, b), 360 (c, d), and $q_{\rm in}=$ 2310 vehicles/h. Other model parameters are the same as those in figure~\ref{Diagram_ATD}.
\label{ATD_J_Pert} } 
\end{center}
\end{figure*}

In the ATD-model, dynamic merging of vehicles from the on-ramp lane onto the main road can cause
dynamic   speed and density waves that propagate upstream in synchronized flow of the SP (figure~\ref{ATD_J_Pert}).
It turns out that
if the flow rate $q_{\rm on}$ is related to a point $(q_{\rm on}, \ q_{\rm in})$ between the boundaries 
$F^{\rm (B)}_{\rm S}$ and $S^{\rm (B)}_{\rm J}$ (figure~\ref{Diagram_ATD} (a)), then these dynamic 
waves decay during their upstream propagation within   synchronized flow of the SP
(figures~\ref{ATD_J_Pert} (a) and (b)). In contrast, at the boundary $S^{\rm (B)}_{\rm J}$
the waves begin to self-growth in their amplitude leading wide moving jam formation, i.e., one of 
  GPs appears upstream of the bottleneck  (figures~\ref{ATD_J_Pert} (c) and (d)).

  As in the KKW cellular automata (CA) model~\cite{KKW,KernerBook}, in the ATD-model
    the maximum flow rate in free flow downstream of the bottleneck
   $q^{\rm (free \ B)}_{\rm max}$ is a decreasing function of $q_{\rm on}$  (figure~\ref{Diagram_ATD} (b)). 
   Recall, that the flow rate $q^{\rm (free \ B)}_{\rm max}(q_{\rm on})$
 is the flow rate in free flow downstream of the bottleneck
 associated with  the boundary $F^{\rm (B)}_{\rm S}$.
  After a congested pattern is formed at the bottleneck, the flow rate downstream of the congested bottleneck
   called discharge flow rate $q^{\rm (bottle)}_{\rm out}$ (figure~\ref{Diagram_ATD} (b)) is usually smaller 
   than the initial flow rate $q^{\rm (free \ B)}_{\rm max}$. The difference  
   $\delta q(q_{\rm on}) =q^{\rm (free \ B)}_{\rm max}(q_{\rm on})-q^{\rm (bottle)}_{\rm out}(q_{\rm on})$ 
 called $\lq\lq$capacity drop"  is an increasing function of
  $q_{\rm on}$ at the boundary $F^{\rm (B)}_{\rm S}$ in the diagram of congested patterns.
  
  In accordance with empirical
results~\cite{KernerBook},  
 in the ATD-model moving jams do {\it not} emerge spontaneously in free flow. This is because in all states of free flow
 critical perturbations required for an
F$\rightarrow$S transition are considerably smaller than those for F$\rightarrow$J transition. 
In the model, all
synchronized flow states that are above the line $J$ in the flow--density plane
(figure~\ref{Steady_states} (c)) are metastable ones against wide moving jam emergence.

\section{Phase Transitions and Congested Patterns in SA-Models \label{SA_S}}
  
  \subsection{Nucleation and Metastability Effects of Pattern Formation \label{ATD_SA_Meta}}
   
\begin{figure*}
\begin{center}
\includegraphics[width=14 cm]{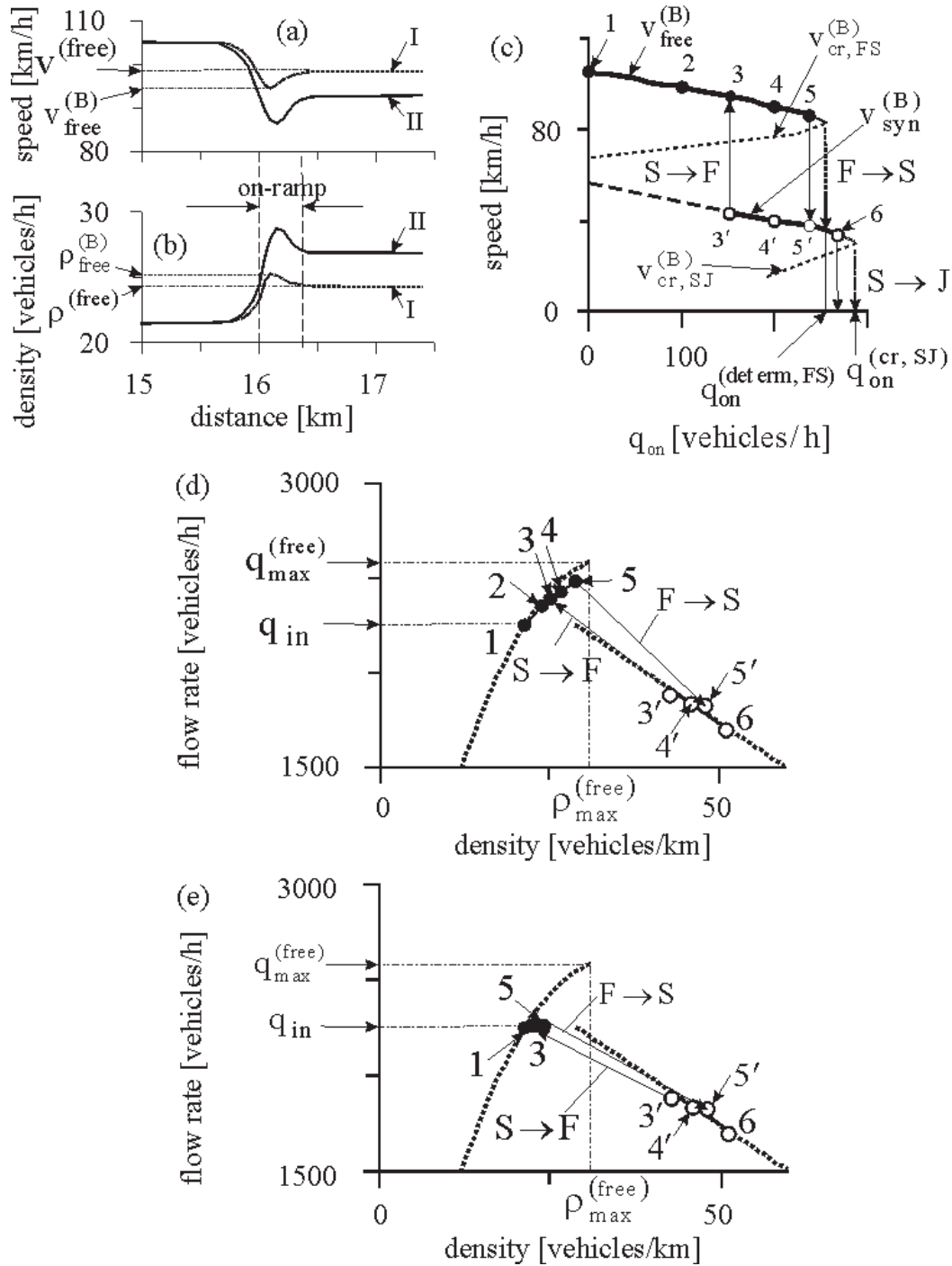}
\caption{Nucleation and metastability of free flow and synchronized flow:
(a, b) -- Spatial dependence of the speed (a) and density (b)
within the deterministic perturbation in free flow at the bottleneck for two flow rates
$q_{\rm on}=$ 150 (curve I) and 240 (curve II) vehicles/h. (c) --
Double Z-characteristic in the flow-rate--speed plane 
for the sequence of F$\rightarrow$S$\rightarrow$J transitions.  (d, e) --
Hysteresis effects in the flow--density plane due to F$\rightarrow$S and reverse S$\rightarrow$F transitions
within the deterministic perturbation  (d) and on the main road upstream of the bottleneck (e).
Dotted curves in (d, e) are related to steady state model solutions. 
$q_{\rm in}=$ 2252 vehicles/h.
\label{SA_Pert} } 
\end{center}
\end{figure*}

\begin{figure*}
\begin{center}
\includegraphics[width=12 cm]{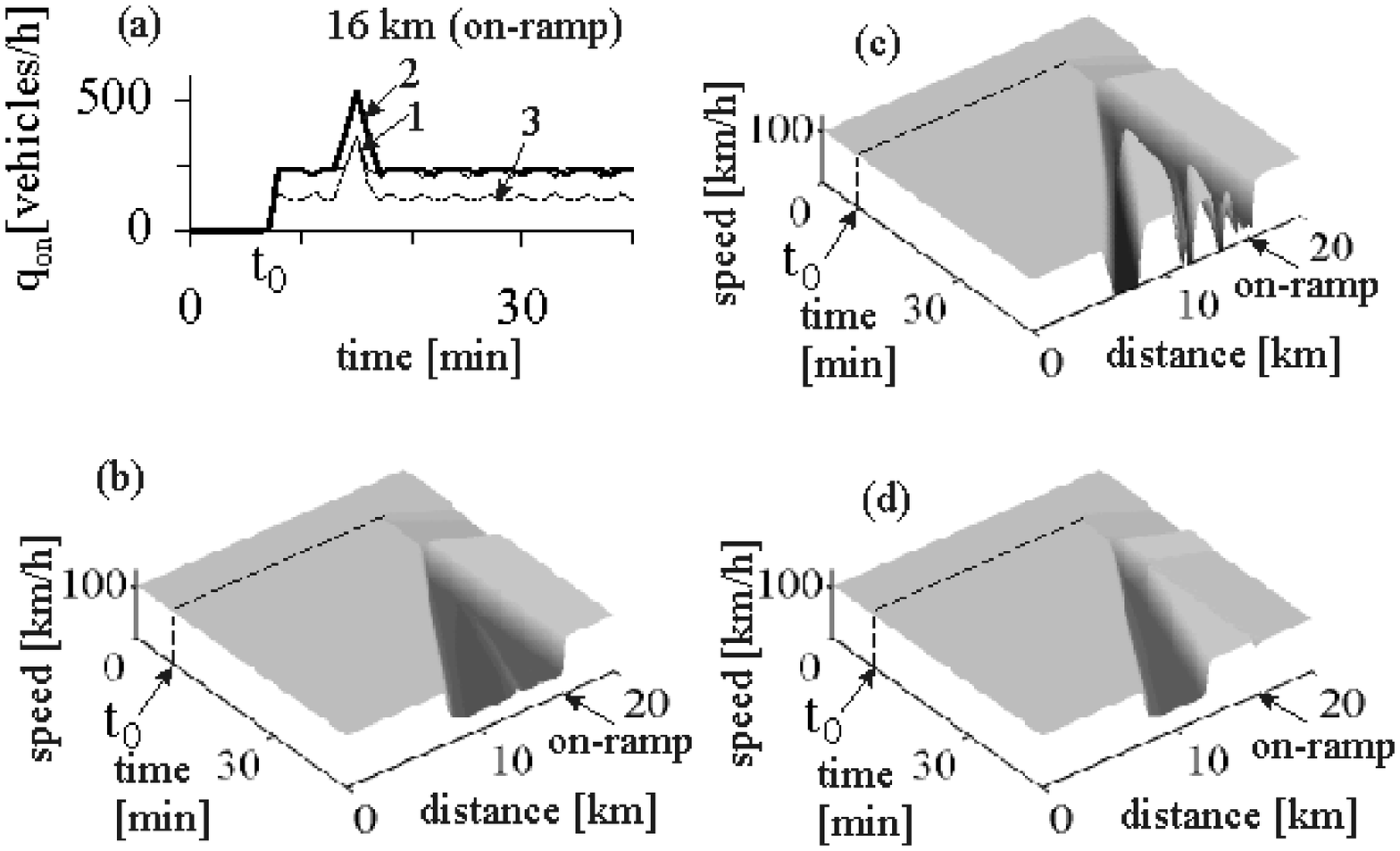}
\caption{Congested pattern excitation in
metastable free flow: (a) -- Short-time (2 min) perturbations in the flow rate $q_{\rm on}$
used for pattern excitation. (b--d) -- WSP (b), GP (c), and MSP (d) induced by the related perturbations
(curve 1, 2, and 3 in (a), respectively).  
$q_{\rm in}=$ 2252 vehicles/h.
$q_{\rm on}=$ 220 vehicles/h in (b, c) and   $q_{\rm on}=$ 140 vehicles/h in (d).
Amplitudes of   perturbations in (a) are:
 270  for (b, d) and  500 vehicles/h for (c).
\label{SA_Meta} } 
\end{center}
\end{figure*}

As the ATD-model, the SA-models exhibit a first-order F$\rightarrow$S transition at the bottleneck, which is accompanied by
 nucleation and metastability effects, as well as by a hysteresis in SP emergence and dissolution.
To illustrate these  effects   found for both the ATD- and SA-models, 
we restrict a consideration to the SA-model (\ref{coor_sa})--(\ref{v_syn_sa_2}) 
(figures~\ref{SA_Pert} and~\ref{SA_Meta}). 
 When an initial state at the bottleneck is free flow in which $q_{\rm in}$ is given and  $q_{\rm on}$
  increases gradually,
 then, as in the ATD-model (figure~\ref{ATD_Pert} (a)),
  a dynamic disturbance in free flow localized at the bottleneck appears spontaneously.
A time averaging of spatial speed and density distributions    within the perturbation leads to the associated deterministic perturbation 
 (figures~\ref{SA_Pert} (a) and (b)). Deterministic perturbation features are the same as those for the ATD-model 
(Sect.~\ref{ATD_LP_Sect}).

\begin{figure*}
\begin{center}
\includegraphics[width=10 cm]{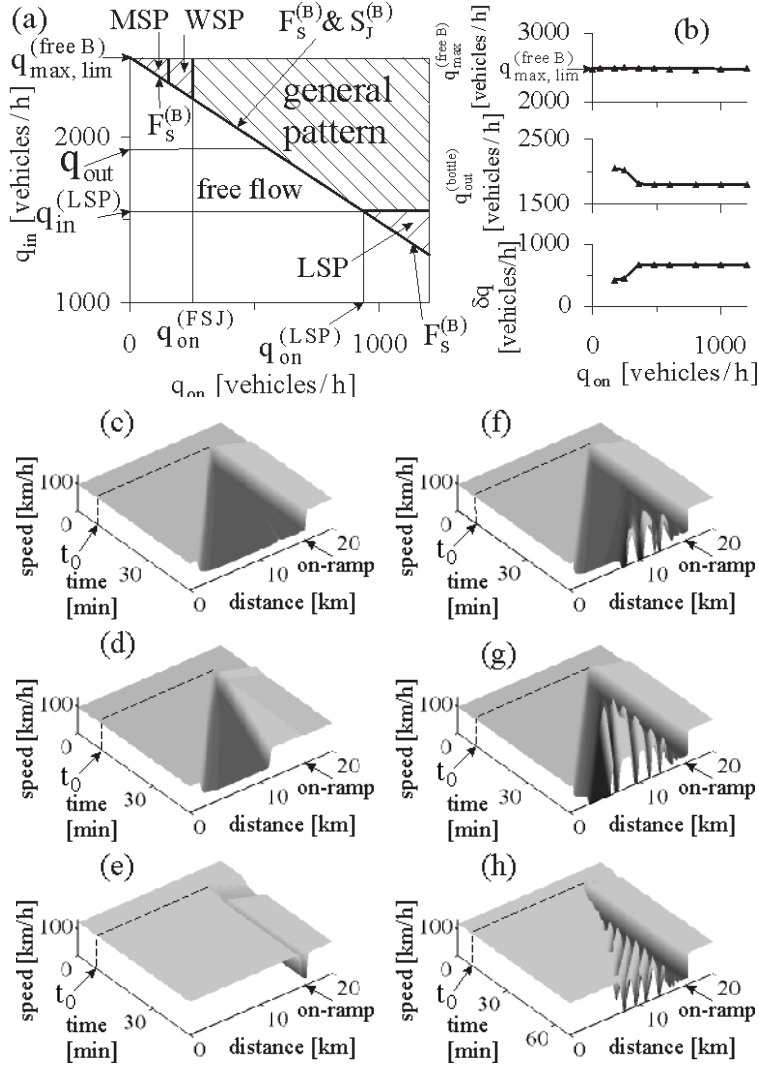}
\caption{Diagram of congested patterns at the on-ramp bottleneck in SA-model (\ref{coor_sa})--(\ref{v_syn_sa_2})
(a),  the maximum capacity in
free flow $q^{\rm (free \ B)}_{\rm max} $,
 the discharge flow rate at the on-ramp $q^{\rm (bottle)}_{\rm out}$
 and the capacity drop $\delta q$  (b), and congested patterns (c--h) related to (a):
(c--e) -- SPs and (f--h) -- GPs. 
(c) -- WSP.
(d) -- MSP.
(e) -- LSP.
(f) -- GP arising from WSP at smaller $q_{\rm  on}$.
(g) -- GP at $q_{\rm  in}>q_{\rm  out}$. 
(h) -  GP at $q_{\rm  in}<q_{\rm  out}$.
In (c--h) 
the flow rates 
$(q_{\rm  on}, q_{\rm  in})$ are: 
(c) (200, 2353), 
(d) (140, 2378), 
(e) (940, 1520),
(f) (250, 2353), (g) (400, 2353), and (h) (900, 1770) vehicles/h. 
$q^{\rm (free \ B)}_{\rm max, \ lim}\approx 2475 $ vehicles/h.
\label{Diagram_SA} } 
\end{center}
\end{figure*}

The  speed $v^{\rm (B)}_{\rm free}$  within the deterministic perturbation   decreases when $q_{\rm on}$ increases
(from point 1 to 5 in figure~\ref{SA_Pert} (c)). Consequently,  $\rho^{\rm (B)}_{\rm free}$ increases.
In the flow--density plane, the flow rate on the main road associated with this density
 increases too (from point 1 to 5 in figure~\ref{SA_Pert} (d)), whereas the flow rate upstream of the perturbation
 is equal to $q_{\rm in}$, i.e., it does not change (from point 1 to 5 in figure~\ref{SA_Pert} (e)).
This increase in the deterministic perturbation amplitude with $q_{\rm on}$
has a limit $q_{\rm on}=q^{\rm (determ, \ FS)}_{\rm on}$ associated with the deterministic F$\rightarrow$S
transition (dotted down-arrow in figure~\ref{SA_Pert} (c)). However, non-homogeneous free flow dynamics (Sect.~\ref{ATD_LP_Sect}), which is   caused by  vehicle
 merging, results in an F$\rightarrow$S
transition at a smaller  $q_{\rm on}$ (point 5 in figure~\ref{SA_Pert} (c))
 related to a point on the boundary $F^{\rm (B)}_{S}$ in the diagram 
of congested patterns (figure~\ref{Diagram_SA} (a)).
The speed decreases and density increases abruptly within the initial perturbation
(arrows F$\rightarrow$S from point 5 to $5^{\prime}$ in figures~\ref{SA_Pert} (c)--(e)) and a congested pattern emerges at the bottleneck. 
In the example,
a widening SP (WSP) occurs upstream of the bottleneck due to the F$\rightarrow$S transition
(figure~\ref{Diagram_SA} (c)).

If now  $q_{\rm on}$ decreases, the speed within the WSP increases
(from point $5^{\prime}$ to $3^{\prime}$ in figures~\ref{SA_Pert} (c)--(e)).
This synchronized flow speed increase  has a limit:
The speed increases and density decreases abruptly within the synchronized flow    
(arrows S$\rightarrow$F from the point $3^{\prime}$ to 3 in figures~\ref{SA_Pert} (c)--(e)) and free flow returns at the bottleneck.

Note that $q_{\rm in}$ in figure~\ref{SA_Pert} is chosen to be greater than the threshold flow rate $q_{\rm th}$ for moving
SP (MSP) existence. 
As a result, the initial motionless downstream front of synchronized flow at the bottleneck begins to move away upstream.
Consequently, an MSP emerges (figure~\ref{Diagram_SA} (d)) (range of $q_{\rm on}$ within which MSPs occur is shown by
a dashed part of the synchronized flow states $v^{\rm (B)}_{\rm syn}$ in figure~\ref{SA_Pert} (c)). 
At greater $q_{\rm on}$ on
the dashed part of the synchronized flow states $v^{\rm (B)}_{\rm syn}$ in figure~\ref{SA_Pert} (c)
this free flow at the bottleneck can persist for a short time only:
A new F$\rightarrow$S transition occurs spontaneously and a new MSP emerges at the bottleneck, and so on.
 Due to this effect, a sequence of MSPs
appears.  

At $q_{\rm in}>q_{\rm th}$ an MSP can also be induced by application  a short-time local perturbation in free flow. 
The speed within this external perturbation should be lower than the critical speed $v^{\rm (B)}_{\rm cr, \ FS}$
associated with the critical branch on the Z-characteristic for the
 F$\rightarrow$S and reverse S$\rightarrow$F transitions at the bottleneck. As in the stochastic models~\cite{KernerBook},
this Z-characteristic  
  consists of the states for  free flow associated with the deterministic perturbation at the bottleneck
$v^{\rm (B)}_{\rm free}$, the critical branch 
$v^{\rm (B)}_{\rm cr, \ FS}$, and synchronized flow states $v^{\rm (B)}_{\rm syn}$ (figure~\ref{SA_Pert} (c)).
In accordance with this Z-characteristic, we get the associated hysteresis effects
 on the fundamental diagram (arrows F$\rightarrow$S and S$\rightarrow$F 
  in figures~\ref{SA_Pert} (d) and (e)).

If in contrast $q_{\rm on}$ increases, the speed within the WSP decreases
(figure~\ref{SA_Pert} (c)). This speed decrease has a limit associated with the flow rate
$q_{\rm on}=q^{\rm (cr, \ SJ)}_{\rm on}$ at which an
 S$\rightarrow$J  transition must occur (dotted down-arrow S$\rightarrow$J in figure~\ref{SA_Pert} (c)). However, because there are speed and density 
 waves of a finite amplitude  in synchronized flow, an S$\rightarrow$J  transition
 occurs already for $q_{\rm on}<q^{\rm (cr, \ SJ)}_{\rm on}$
 (point 6 and solid down-arrow  in figure~\ref{SA_Pert} (c)). As a result, an GP emerges. 
This is because in the SA-models, 
 synchronized flow steady states with the speed $v> v^{\rm (SJ)}_{\rm cr}$, which 
  are above the line $J$ in the flow--density plane,
 are metastable ones against wide moving jam emergence.
This metastability can be seen from another Z-characteristic in the speed--flow plane
associated with an
 S$\rightarrow$J  transition in synchronized flow. The Z-characteristic consists of the states for synchronized flow 
$v^{\rm (B)}_{\rm syn}$, the critical branch for critical perturbations in synchronized flow
$v^{\rm (B)}_{\rm cr, \ SJ}$, and the line $v=0$ for wide moving jams (figure~\ref{SA_Pert} (c)).

From the resulting double Z-characteristic (figure~\ref{SA_Pert} (c)), it can concluded that in a metastable free flow at the bottleneck
(left of the boundary $F^{\rm (B)}_{\rm S}$ in the diagram in figure~\ref{Diagram_SA} (a))
depending on amplitude of a time-limited perturbation caused, for example, by an increase in  
$q_{\rm on}$ (curves 1 and 2 in figure~\ref{SA_Meta} (a)), either an WSP (figure~\ref{SA_Meta} (b)) or an GP (figure~\ref{SA_Meta} (c))
can be induced. At smaller  $q_{\rm on}$ (curve 3 in figure~\ref{SA_Meta} (a)), an MSP (figure~\ref{SA_Meta} (d))
can be excited in free flow. All  results presented in
figures~\ref{SA_Pert} and~\ref{SA_Meta} for the SA-model remain   qualitatively equal  for the ATD-model.

\subsection{Comparison of Congested Patterns in ATD- and SA-Models \label{SA_Con_Sec}}

The SA-model (\ref{acc_sa})--(\ref{v_syn_sa_2}) (figure~\ref{Diagram_SA})
exhibits the following shortcoming   in comparison with
the ATD-model (figure~\ref{Diagram_ATD}):

(i) If the flow rate $q_{\rm on}$ is within a flow rate range $q^{\rm (FSJ)}_{\rm on}<q_{\rm on}<q^{\rm (LSP)}_{\rm on}$,
then no SP can be formed at the boundary $F^{\rm (B)}_{\rm S}$ in the diagram (figure~\ref{Diagram_SA} (a)): The sequence of 
  F$\rightarrow$S$\rightarrow$J transitions occurs spontaneously at this boundary, leading to 
 GP emergence. For this reason, the related part of the boundary  at which GPs
 emerge spontaneously in free flow at the bottleneck is labelled $F^{\rm (B)}_{\rm S} \ \& \ S^{\rm (B)}_{\rm J}$.
 
 (ii) If the flow rate $q_{\rm in}$ at this boundary decreases, another
 characteristic flow rate $q_{\rm in}=q^{\rm (LSP)}_{\rm in}$ associated with the flow rate
 $q_{\rm on}=q^{\rm (LSP)}_{\rm on}$ at this boundary is reached:
 At $q_{\rm in}<q^{\rm (LSP)}_{\rm in}$    moving jams do not emerge in synchronized flow
 upstream of the bottleneck. As a result, at $q_{\rm in}<q^{\rm (LSP)}_{\rm in}$ and right of 
  the boundary $F^{\rm (B)}_{\rm S}$ only an LSP remains at the bottleneck. Within this LSP the speed is very low.
  This LSP has a qualitative
  different nature in comparison with an LSP of higher synchronized flow 
  speed in    the ATD-model that occurs at considerably greater $q_{\rm in}$
 (figure~\ref{Diagram_ATD}).
 
\begin{figure*}
\begin{center}
\includegraphics[width=10 cm]{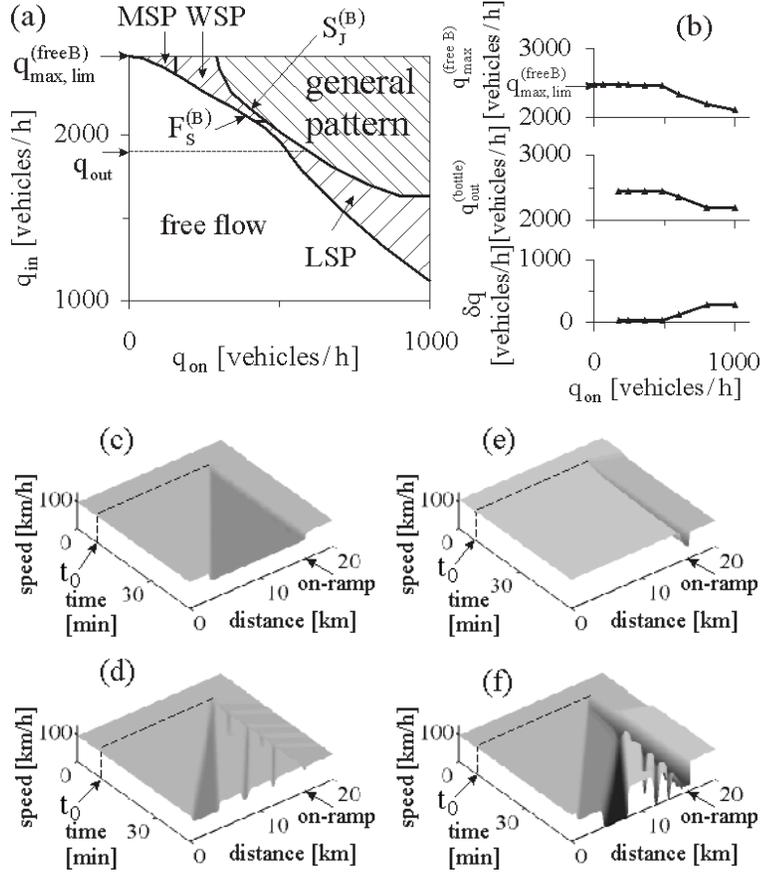}
\caption{Diagram of congested patterns at the on-ramp bottleneck in the SA-model (\ref{coor_sa})--(\ref{a_jam_sa}),
 (\ref{v_syn_sa_1}) (a), the
maximum  capacity in
free flow $q^{\rm (free \ B)}_{\rm max} $,
 the discharge flow rate at the on-ramp $q^{\rm (bottle)}_{\rm out}$
 and the capacity drop $\delta q$ (b),
and congested patterns (c-f) related to (a):
(c-e) - SP and (f) - GP. 
(c) - WSP.
(d) - Subsequence of MSPs.
(e) - LSP.
(f) - GP arising from WSP.
In (c-f), 
the flow rates 
$(q_{\rm  on}, q_{\rm  in})$ are: 
(c) (200, 2400), 
(d) (90, 2450), 
(e) (360, 2115),
(f) (300, 2400) vehicles/h. 
$q^{\rm (free \ B)}_{\rm max, \ lim}\approx 2475 $ vehicles/h.
In (b),  $q^{\rm (bottle)}_{\rm out}$ is changed
from 2450   to 2200  vehicles/h. 
$q_{\rm out} \approx 1880$ vehicles/h. 
\label{Diagram_SA_positive} } 
\end{center}
\end{figure*}

\begin{figure*}
\begin{center}
\includegraphics[width=10 cm]{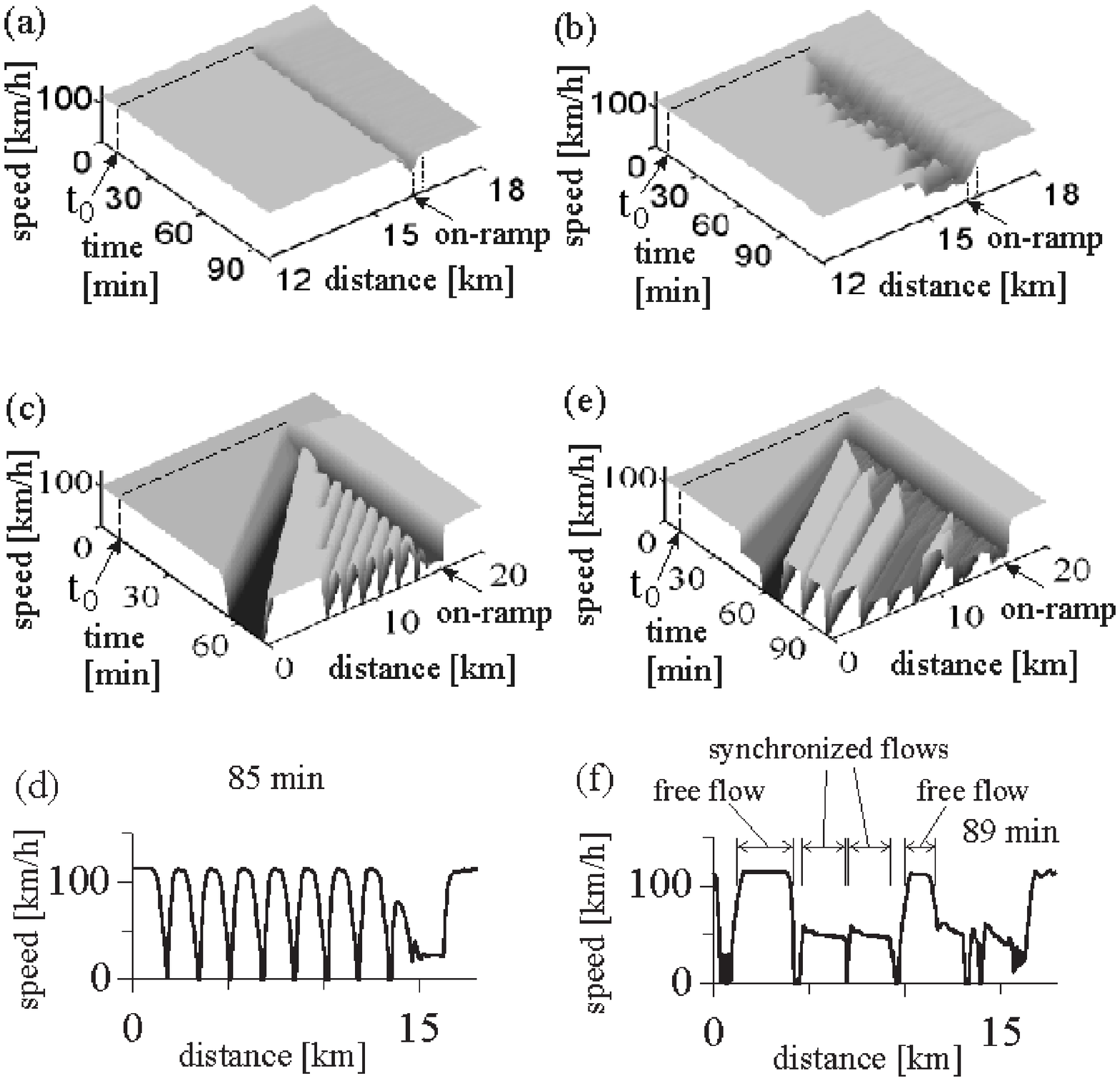}
\caption{Comparison of LSPs (a, b)  and GPs (c--f) in the SA-model (\ref{coor_sa})--(\ref{a_jam_sa}),
 (\ref{v_syn_sa_1}) (a, c, d) and in the ATD-model  (b, e, f). 
LSPs and their parameters are the same as those in figures~\ref{Diagram_SA_positive} (e) and~\ref{Diagram_ATD} (e), respectively.
In (e, f), $\Delta v^{(1)}_{\rm r}=$ 7.5 m/s. $(q_{\rm on}, \  q_{\rm in})$ are: (c, d) (380, \ 2397), (e, f) (660, \ 2222)
 vehicles/h.
\label{LSP_ATD_SA} } 
\end{center}
\end{figure*}

 In the SA-model (\ref{coor_sa})--(\ref{a_jam_sa}), (\ref{v_syn_sa_1}),
  the branch for average synchronized flow states $V_{\rm av}^{\rm (syn)}$   has a part with a positive slope 
 (figure~\ref{Steady_states_SA} (d)). Then LSPs of higher speeds appear in the diagram of congested patterns (figure~\ref{Diagram_SA_positive}).
 However, these LSPs are not related to LSPs observed in empirical 
 observations.
 To explain this, note that these model LSPs are very narrow ones (figure~\ref{Diagram_SA_positive} (e)). 
 They are localized within the merging region of the on-ramp and
 consist of two narrow fronts only (figure~\ref{LSP_ATD_SA} (a)): There is no    region of synchronized flow between the fronts within these LSPs.
 This is regardless of the flow rates $q_{\rm in}$ and $q_{\rm on}$.
  Conflictingly, in empirical observations
 rather than such narrow LSPs, an extended region of synchronized flow is usually observed within an empirical LSP. 
 The LSP width (in the longitudinal direction) changes over time considerably. These empirical features of LSPs  shown by the ATD-model
 (figure~\ref{LSP_ATD_SA} (b)) are not found in the SA-model.

  (iii) In the ATD-model (figures~\ref{LSP_ATD_SA} (e) and (f)) as in empirical observations,
  both free and synchronized flows can be formed between wide moving jams within an GP.
 In contrast,   in the SA-models only  free flow
  can  be formed between wide moving jams within the GP
     (figures~\ref{LSP_ATD_SA} (c) and (d)). The reason for this is as follows. 
     The average branch for synchronized flow   lies for  speeds $v > v^{\rm (SJ)}_{\rm cr}$
   above the line $J$ (figure~\ref{Steady_states_SA} (g)). Flow states in the jam outflow should be related to points on the line $J$.
   Thus, there are no synchronized flow states between the jams. This explains why only free flow can be formed between the jams in 
   the SA-models.\footnote{The only exclusion is the SA-model with the average branch for synchronized flow (\ref{v_syn_sa_1}),
  if parameters for the curve $S$ and/or the line $J$, i.e., for
   wide moving jam propagation are chosen   different  as those shown 
   in figures~\ref{Steady_states_SA} (c), (d), and (h): These different parameters should lead to an intersection 
   of the line $J$ with
the average branch for synchronized flow with a positive slope in the flow--density plane.
However, in this specific case only one state of the synchronized flow, which is associated with the point of the latter intersection,
is possible. This model effect is not agreed with empirical results, in which the flow rate and speed between wide moving jams within GPs can change
over time considerably~\cite{KernerBook}.}

   The mentioned shortcoming of the SA-model   result from
   the averaging of a 2D-region of steady states for synchronized flow in the flow--density plane
    of the ATD-model (figure~\ref{Steady_states} (a)) to the branch for average synchronized flow states (curve $S$ in
      figure~\ref{Steady_states_SA}).

At   chosen SA-model parameters the condition 
\begin{equation}
q_{\rm out}>q^{\rm (pinch)}
\label{qcr_qout}
\end{equation}
is satisfied, where $q^{\rm (pinch)}$ is the  flow rate within the pinch region of an GP  in which narrow moving jams emerge.
 Under the condition  (\ref{qcr_qout}),
no DGPs appear in the SA-models (figures~\ref{Diagram_SA}  and~\ref{Diagram_SA_positive}).
At other   parameters of the SA-models, an opposite condition
\begin{equation}
q_{\rm out}<q^{\rm (pinch)}
\label{qout_qcr}
\end{equation}
can be satisfied. Then DGPs   appear in the SA-models.

The maximum flow rate in free flow downstream of the bottleneck
   $q^{\rm (free \ B)}_{\rm max}(q_{\rm on})$,  the discharge flow rate $q^{\rm (bottle)}_{\rm out}$,  and
   the $\lq\lq$capacity drop" $\delta q$ can sometimes exhibit different features as those in the ATD-model
    (figure~\ref{Diagram_ATD} (b)) when the flow rate $q_{\rm on}$
 changes  (figures~\ref{Diagram_SA} (b) and~\ref{Diagram_SA_positive} (b)). Particularly, 
 in contrast with the ATD-model, in the SA-model (\ref{coor_sa})--(\ref{v_syn_sa_2}) 
 $q^{\rm (free \ B)}_{\rm max}$ does not depend on $q_{\rm on}$, whereas
 in the SA-model (\ref{coor_sa})--(\ref{a_jam_sa}),
 (\ref{v_syn_sa_1}) $q^{\rm (free \ B)}_{\rm max}$  depends on $q_{\rm on}$ but at
 considerably greater $q_{\rm on}$
 than for the ATD-model.

 Simulations show that the  SA models presented in~\ref{B} show qualitatively the same features of the 
phase transitions and spatiotemporal congested patterns as those in the SA-model (\ref{coor_sa})--(\ref{v_syn_sa_2}).

\subsection{Comparison with Stochastic SA-Models}

It is interesting to compare
the deterministic SA-models  with
 possible {\it stochastic} SA-models. Such models can be derived from the stochastic model of Ref.~\cite{KKl,KKl2003A}, if  
2D region of synchronized flow steady states is averaged to   synchronized flow states related to a 1D region
in the flow--density plane. 

A stochastic SA-model can easily be derived from the stochastic model of Ref.~\cite{KKl2003A}
based on the physics and ideas for the SA-model approach discussed in Sect.~\ref{SA_Section}.
To reach this goal, in the  part of the stochastic model of Ref.~\cite{KKl2003A}
\begin{eqnarray}
\label{next}
v_{\rm n+1}=\max(0, \min({v_{\rm free}, v_{\rm c,n}, v_{\rm s,n} })), \\
x_{\rm n+1}= x_{\rm n}+v_{\rm n+1}\tau  
\end{eqnarray}
for a desired speed in synchronized flow $v_{\rm c,n}$, rather than the formula (3) of Ref.~\cite{KKl2003A}
leading to a 2D region of synchronized flow steady states in the flow--density plane,
 the following equations associated with the physics of the SA-models
of Sect.~\ref{SA_Section} are used:
\begin{eqnarray}
\label{next1}
v_{\rm c,n}= v_{\rm n} +  
\max(-b_{\rm n}\tau, \ \min(a_{\rm n}\tau, \Delta_{\rm n})), 
\end{eqnarray}
\begin{equation}
\Delta_{\rm n}=\left\{
\begin{array}{ll}
A^{\rm (free)}(g_{\rm n})(v_{\rm free}-v_{\rm n})+ \nonumber \\
K(v_{\rm \ell,n}-v_{\rm n}) &  \textrm{at $g \geq g^{\rm (free)}_{\rm min}$ },  \\
A^{\rm (syn)}(V^{\rm (syn)}_{\rm av}(g_{\rm n})-v_{\rm n})   \\
+ K(v_{\rm \ell,n}-v_{\rm n}) &  \textrm{at $ g<g^{\rm (free)}_{\rm min}$}.
\end{array} \right.
\label{next2} 
\end{equation}
In (\ref{next})--(\ref{next2}),  
$v_{\rm n}$ and $x_{\rm n}$ are the speed and 
 space co-ordinate of  a vehicle; the index $n$ corresponds 
to the discrete time $t=n\tau$, $n=0,1,2,..$; $\tau$ is the time step; 
$v_{\rm free}$ is the maximum speed 
in free flow, which is  a constant;
 $v_{\rm s,n}$ is the
save speed of Ref.~\cite{KKl2003A};  
 $a_{\rm n}\geq 0$ is  acceleration, $b_{\rm n}\geq 0$ is  deceleration,
 which
  are taken as the same stochastic functions used in the model of Ref.~\cite{KKl2003A};
the space gap 
$g_{\rm n}=x_{\rm \ell,n}-x_{\rm n}-d$;
the average speed in synchronized flow steady states  
$V^{\rm (syn)}_{\rm av}$ is given by the formula (\ref{v_syn_sa_2}) at $g^{\rm (jam)}_{\rm max}=0$.
 Of course, other formulations for the average synchronized flow steady states
$V^{\rm (syn)}_{\rm av}$, for example used in
the deterministic SA-models (figures~\ref{Steady_states_SA} (b), (d), and (f)) can also be applied.

In general, descriptions of   random vehicle acceleration and deceleration 
are the same as those in the stochastic model of Ref.~\cite{KKl,KKl2003A}: At the first step,
the preliminary speed $\tilde v_{\rm n+1}$ is set to $\tilde v_{\rm n+1}=v_{\rm n+1}$
where the speed $v_{\rm n+1}$ is calculated from the  equations (\ref{next})--(\ref{next2}).
At the second step, a noise component $\xi_{\rm n}$ is added to the calculated speed $\tilde v_{\rm n+1}$. Then
the final speed is found from the condition~\cite{KKl2003A}
\begin{equation}
v_{\rm n+1}=\max(0, \min({v_{\rm free}, \tilde v_{\rm n+1}+\xi_{\rm n}, v_{\rm n}+a_{\rm max} \tau, v_{\rm s,n} })),
\label{final}
\end{equation}
where $a_{\rm max}$ is the maximum acceleration.

However, in contrast with the stochastic model of Ref.~\cite{KKl2003A},
the noise component $\xi_{\rm n}$ in (\ref{final})  
is chosen to be different from zero {\it only} if the vehicle 
decelerates, specifically
\begin{equation} 
\xi_{\rm n}=\left\{
\begin{array}{ll}
-b_{\rm max}\tau \theta(p_{\rm b}- r) &  \textrm{if $\tilde v_{\rm n+1}< v_{\rm n}-\delta$} \\
0 &  \textrm{otherwise},
\end{array} \right.
\label{noise}
\end{equation} 
where $r= {\rm rand}(0,1)$,
$\theta (z)=0$ at $z<0$ and $\theta (z)=1$ at $z \geq 0$,
$\delta\ll \tau a_{\rm max}$, $b_{\rm max}$,  $p_{\rm b}$ are constants.

\begin{figure*}
\begin{center}
\includegraphics[width=10 cm]{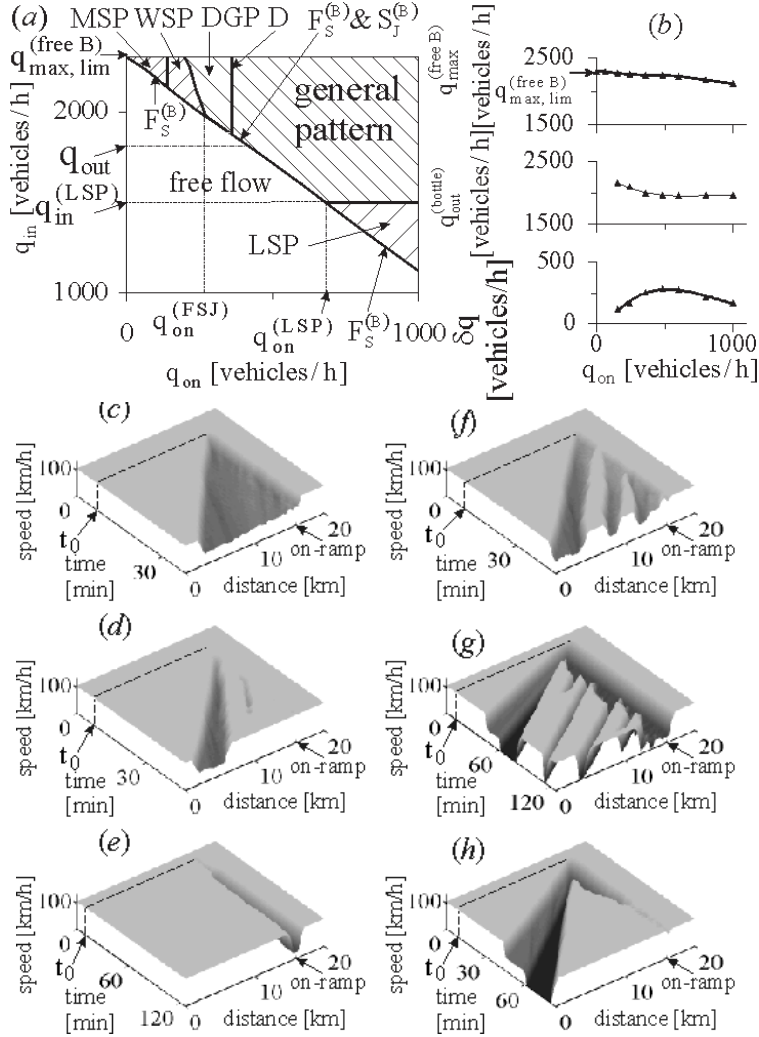}
\caption{Diagram of congested patterns at the on-ramp bottleneck in the stochastic SA-model (\ref{next})--(\ref{noise})
(a), the maximum capacity in
free flow $q^{\rm (free \ B)}_{\rm max} $,
 the discharge flow rate at the on-ramp $q^{\rm (bottle)}_{\rm out}$
 and the capacity drop $\delta q$ (b), and congested patterns (c--h) related to (a):
(c--f) -- SPs and (g, h) -- GPs. 
(c) -- WSP.
(d) -- MSP.
(e) -- LSP.
(f) -- ASP.
(g) -- GP. 
(h) -  DGP.
In (c--h) 
the flow rates 
$(q_{\rm  on}, q_{\rm  in})$ are: 
(c) (200, 2250), 
(d) (60, 2250), 
(e) (720, 1470),
(f) (120, 2235), 
(g) (500, 2220),
and (h) (352, 2235) vehicles/h. 
$q^{\rm (free \ B)}_{\rm max, \ lim}\approx$ 2300 vehicles/h. Model parameters:
$A^{\rm (free)}(g_{\rm n})=0.5 \min \big(1, (g_{\rm n}-g^{\rm (free)}_{\rm min})/20 \big)$,
 $g^{\rm (free)}_{\rm min}=$ 36 m,
$A^{\rm (syn)}=0.1$,
 $T^{\rm (syn)}_{\rm av}=1.45$ s, $g^{\rm (jam)}_{\rm max}=$ 0,
$p_{\rm b}=$ 0.02. $p_{1}=$0.55, $p_{2}(v_{\rm n})=0.5+0.48\theta(v_{\rm n}-15)$, $a_{\rm max}=b_{\rm max}=$ 0.5 $\rm m/s^{2}$,
$K=$ 1.
 The other parameters are the same as those in~\cite{KKl2003A}. The
on-ramp model from~\cite{KKl2003A, KKl2004AA} is used, in which the synchronization gaps
$G^{+}_{n}=G^{-}_{n}=g^{\rm (free)}_{\rm min}$.
\label{Diagram_SA_S} } 
\end{center}
\end{figure*}

Simulations show that the stochastic SA-model (\ref{next})--(\ref{noise}) exhibits qualitatively similar  spatiotemporal congested   patterns at the on-ramp bottleneck
(figure~\ref{Diagram_SA_S}) as those in
the associated deterministic SA-models (figure~\ref{Diagram_SA}). However, there are
qualitative differences in the dynamics of first-order F$\rightarrow$S and S$\rightarrow$J transitions leading to pattern formation
explained in Sect.~\ref{ATD_SA_Meta}: In  the stochastic SA-model, random model fluctuations
are important for phase transition nucleation, whereas in the deterministic SA-models
the F$\rightarrow$S and S$\rightarrow$J transitions are nucleated by dynamic
perturbations emerging within the on-ramp merging region.

Note that
under the chosen model parameters in the stochastic SA-model (\ref{next})--(\ref{noise})
the condition (\ref{qout_qcr}) can be satisfied at smaller $q_{\rm  on}$. As a result, there is a region in the diagram of congested
patterns in which DGPs occur (region labelled $DGP$ in figure~\ref{Diagram_SA_S} (a)). After the wide moving jam of the DGP is 
upstream of the bottleneck as well as in the ATD-model, an LSP remains at the bottleneck
(figure~\ref{Diagram_SA_S} (h)). In contrast with the ATD-model, this LSP exists for
a finite time interval only and free flow  returns  at the bottleneck. 

\begin{figure*}
\begin{center}
\includegraphics[width=10 cm]{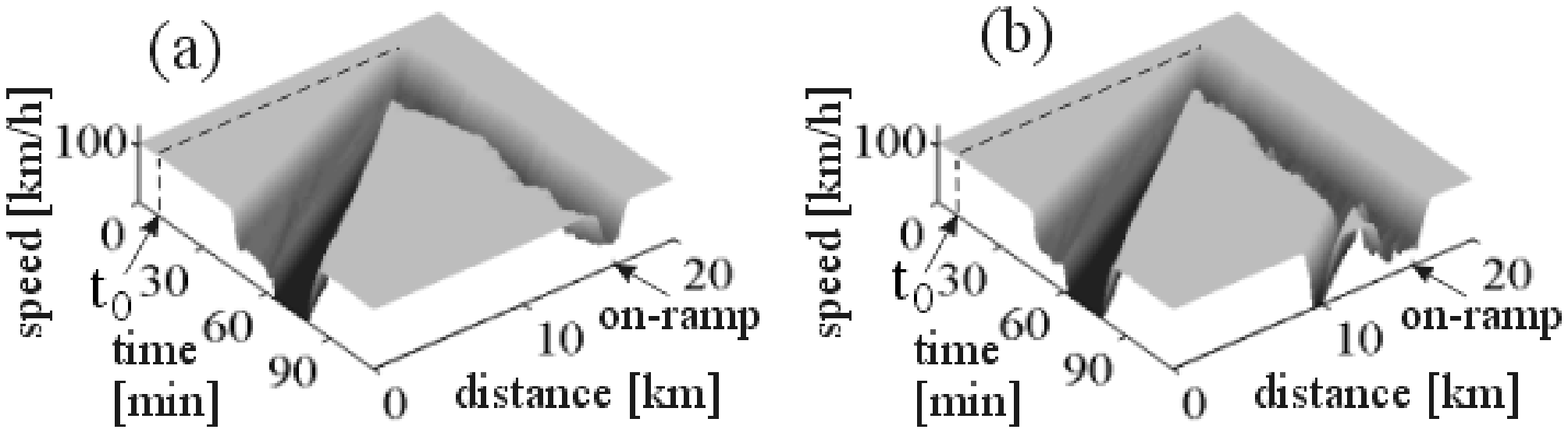}
\caption{Transformation of DGP in   figure~\ref{Diagram_SA_S} (h) to GP  with
 a very long time interval between wide moving jam emergence (b)
through an DGP with a quasi-steady LSP (a) by a small increase
in the flow rate $q_{\rm  on}$ in a neighbourhood of the boundary $D$ in the diagram of congested patterns
in figure~\ref{Diagram_SA_S} (a).
The flow rates 
$(q_{\rm  on}, q_{\rm  in})$ are: 
(a) (357, 2235) and 
(b) (362, 2235) vehicles/h. 
\label{DGP_GP_LSP} } 
\end{center}
\end{figure*}

However, in a small neighbourhood of the boundary
labelled $D$ in the diagram, which separates DGPs and GPs, there is a
peculiarity in pattern formation
 under the condition (\ref{qout_qcr}):  If the flow rate $q_{\rm  on}$ increases in the neighbourhood of the
boundary $D$ in the diagram, then the lifetime of an LSP, which occurs within an DGP
increases and it tends towards infinity at the boundary $D$ (figure~\ref{DGP_GP_LSP} (a)).
This quasi-steady LSP is explained by a very long interval between wide moving jam emergence
in the synchronized flow at the bottleneck (figure~\ref{DGP_GP_LSP} (b)).
This interval tends towards the infinity at the boundary $D$.

\section{Discussion
\label{Discussion}}

\subsection{Comparison of ATD- and SA-models with OV-models and other Deterministic Models \label{Comparison}}

The first term in the formula for vehicle acceleration  $\tilde a^{\rm (free)}$ (\ref{a_free}), $A(V^{\rm (free)}(g)-v)$, describes
the dynamics of the  speed $v$  in the vicinity of the optimal speed  $V^{\rm (free)}(g)$ in free flow.
At a time scale that is considerably greater than the time delay $\tau$, this dynamic behaviour is
the same as those in different OV-models~\cite{Sch,Helbing2001,Nagatani_R,B1995A}, which can be written as follows
\begin{equation}
\frac{dv}{dt}=A(g, \ v)(V(g)-v).
\label{a_OV}
\end{equation}
However, in (\ref{a_OV}) the vehicle acceleration $A(g, \ v)(V(g)-v)$ is valid for the whole possible space gap 
 range~\cite{Sch,Helbing2001,Nagatani_R,B1995A}
\begin{eqnarray} 
g\geq 0.
\end{eqnarray}
In contrast with the OV-models, 
in the ATD-model this vehicle acceleration is applied  for large space gaps
(\ref{G_g}) associated with
free flow {\it only}. 

The crucial difference of the ATD-model with the OV-models and all other deterministic microscopic
traffic flow models (see 
references in the reviews~\cite{Gartner,Sch,Helbing2001,Nagatani_R,Nagel2003A}) is that
the vehicle acceleration  behaviour qualitatively changes when the vehicle is within the synchronization gap, i.e., if
the  condition
(\ref{g_G}),
which is opposite to the condition (\ref{G_g}), is satisfied. 

The condition (\ref{g_G})
is associated with 
the synchronized flow phase in which  there is {\it no} optimal speed in the  ATD-model. This conclusion follows from (\ref{a_syn})
and its analysis made in Sect.~\ref{SteadyStates}  in which it has been shown that
for a given steady space gap in synchronized flow there are an infinity of steady vehicle speeds within a finite speed range
(figure~\ref{Steady_states} (a)).   

The concept of  safe speed $v_{\rm s}(g, \ v_{\ell})$  for vehicle collision prevention
used in the ATD-model
is qualitatively different from the concept of
optimal speed  that is the basis of the deterministic approaches (\ref{Eq_a_opt}) and
 (\ref{Eq_v_opt}): The optimal speed is a desired one (this explains the term $\lq\lq$optimal" speed)
 for a driver to be reached (the driver   moves comfortable with the optimal speed
 during a long time), whereas the safe speed
 is not an optimal one but a limiting speed that is still permitted (the driver should not move with this speed
 during a long time because this is strain for the driver and, therefore, non-comfortable). 
 The qualitative difference of these two concepts
 is mathematically reflected  in the dynamic model behaviour.
 In the ATD-model, when the vehicle speed is higher than the safe speed and
 safe deceleration is applied, then
 a driver time delay   is equal to a small 
 driver reaction time: $\tau=\tau_{\rm s}$ (\ref{tau}).
In all other driving situations,
which are not associated with safe speed, driver time delays are different from $\tau_{\rm s}$. 
This is because  these driver time delays are associated mostly with qualitatively different
expected events occurring within  different traffic phases 
(Sect.~\ref{time_del_sect}). As a result, in the ATD-model driver deceleration to the safe speed
 occurs considerably quicker, then in other driving situations.
 In contrast, in accordance with the concept of optimal speed, in OV-models there is
a   driver time 
 delay in deceleration  that characterizes  speed relaxation to
 the optimal  
 speed~\cite{Gartner,Helbing2001,Nagatani_R,New2,Wh,B1995A,Davis2003B}.

The crucial differences between the  SA-models and    all other   
traffic flow models in which steady states covering a one-dimensional region(s) in the flow--density plane (see 
references in the reviews~\cite{Gartner,Sch,Helbing2001,Nagatani_R,Nagel2003A})
are as follows. In contrast with the models of Ref.~\cite{Gartner,Sch,Helbing2001,Nagatani_R,Nagel2003A}, in the SA-models
 at each density of free flow states  
 the critical amplitude of a local perturbation
required for an F$\rightarrow$S transition is considerably smaller than the critical amplitude of a local perturbation
 required for an F$\rightarrow$J transition.

In the SA-models, there are two  ranges  of
model steady states
  separated one from another by a model discontinuity in vehicle space gap
 or in speed  (figures~\ref{Steady_states_SA} (a)--(e)) or else due to instability of model
 steady states  against infinitesimal non-homogeneous fluctuations (figure~\ref{Steady_states_SA} (f))   
This simulates
the hypothesis of three-phase traffic theory about a competition between over-acceleration and speed adaptation effect that is responsible for 
  F$\rightarrow$S and S$\rightarrow$F transitions: The first range of steady states simulates free flow, whereas
  the second simulates synchronized flow. 
  To simulate  S$\rightarrow$J transitions within synchronized flow, 
steady states associated with synchronized flow of higher speeds  are metastable with respect to
moving jam emergence, i.e., moving jams emerge in these synchronized flow states {\it only} if large enough amplitude
local perturbations appear; synchronized flow states of lower speeds are unstable with respect to moving jam emergence. 
These requirements to the SA-models
lead to F$\rightarrow$S$\rightarrow$J transitions 
 that are responsible for moving jam emergence found in empirical data~\cite{KernerBook}.

\subsection{Critical Discussion of Theories and Models based on the Fundamental Diagram Approach \label{Critical}}

\begin{figure*}
\begin{center}
\includegraphics[width=10 cm]{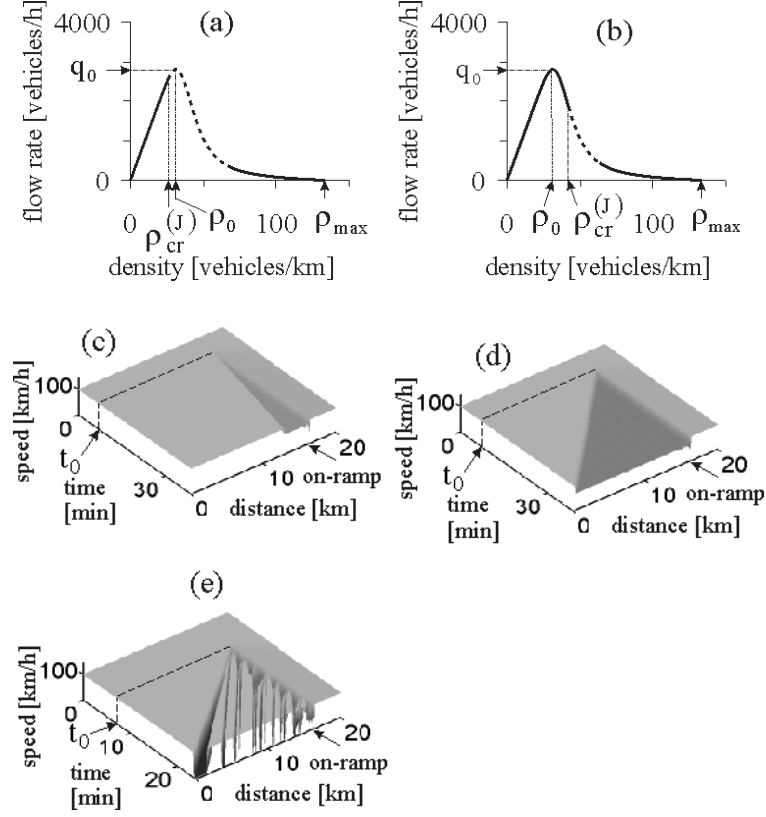}
\caption{Traffic patterns in OV-models at an on-ramp bottleneck:
(a, b) --
Fundamental diagrams of OV-models when the condition (\ref{FJ_01}) (a) and (\ref{FJ_02}) (b) are satisfied, respectively.
(c, d) -- Widening patterns of dense flow upstream of the bottleneck  at two flow rates
$q_{\rm on}=$160 (c) and  400 (d) vehicles/h.
(e) -- 
Formation of wide moving jams within the dense flow upstream of the bottleneck
at $q_{\rm on}=$700 vehicles/h.
In (c--e) the flow rate $q_{\rm in}=$2676 vehicles/h.
Figures (c--e) are related to the OV-model (b) at
$V(g)=V_{0} \big(\tanh((g-g_{0})/g_{1})+\tanh(g_{0}/g_{1}) \big)$ at $V_{0}=$ 14 m/s,
 $g_{0}=$ 17 m,  $g_{1}=$ 7 m, the sensitivity $A(g, \ v)=A(v)$ in (\ref{a_OV})
is $A(v)= \ 5 \ s^{-1}$ at $v \geq $ 12 m/s and $A(v)=0.9 \ s^{-1}$  at $v < $12 m/s.
\label{OV_patterns} } 
\end{center}
\end{figure*}

In the OV model (\ref{a_OV}), as in other deterministic (and stochastic)
traffic flow models in the context of the fundamental diagram approach reviewed in~\cite{Gartner,Sch,Helbing2001,Nagatani_R,Nagel2003A},
which claim to show spontaneous jam emergence,
there is a range of the density on the fundamental diagram in which
steady states on this diagram are unstable against infinitesimal 
perturbations.\footnote{It should   be noted that some of the models based on
 the fundamental diagram approach are not valid
far from equilibrium. It is not simply a matter of a phase transition type that a model exhibits
in steady conditions, but of 
the difficulty of closing equations, which should work in unsteady conditions by relations valid
only in steady uniform conditions.}
This instability leads to wide moving jam emergence in these models both on homogeneous road and at a bottleneck. 
We denote the minimum density of this density range,
in which infinitesimal fluctuations grow, by $\rho^{\rm (J)}_{\rm cr}$ (figures~\ref{OV_patterns} (a) and (b)).

There are two possibilities for the arrangement of the point of this instability 
$(\rho^{\rm (J)}_{\rm cr}, \ q^{\rm (J)}_{\rm cr})$ on the fundamental diagram
in the OV model and other models in the context of the fundamental diagram approach:

(i) The point   $(\rho^{\rm (J)}_{\rm cr}, \ q^{\rm (J)}_{\rm cr})$ lies left of the maximum point of the fundamental diagram
$(\rho_{0}, \ q_{0})$, i.e.,
on the branch of the diagram with a positive slope (figure~\ref{OV_patterns} (a)): 
\begin{equation}
\rho=\rho^{\rm (J)}_{\rm cr}<\rho_{0}
\label{FJ_01}
\end{equation}

(ii)  The point $(\rho^{\rm (J)}_{\rm cr}, \ q^{\rm (J)}_{\rm cr})$ 
lies right of the maximum point of the fundamental diagram $(\rho_{0}, \ q_{0})$, i.e.,
on the branch of the diagram with a negative slope (figure~\ref{OV_patterns} (b)): 
\begin{equation}
\rho=\rho^{\rm (J)}_{\rm cr}>\rho_{0}.
\label{FJ_02}
\end{equation}

Note that in both cases (i) and (ii) all states on the fundamental diagram, in which the density satisfies
the  condition
\begin{equation}
\rho_{\rm min}\leq  \rho <\rho^{\rm (J)}_{\rm cr},
\end{equation}
where $\rho_{\rm min}$ is the density in the wide moving jam outflow associated with the flow rate
$q_{\rm out}$, are metastable states with respect to moving jam emergence~\cite{KK1994}.
The case (i) has intensively been considered in the literature~\cite{Sch,Helbing2001,Nagatani_R,Nagel2003A}
and criticized in Sect. 3.3.2 of the book~\cite{KernerBook}.

\begin{figure*}
\begin{center}
\includegraphics[width=12 cm]{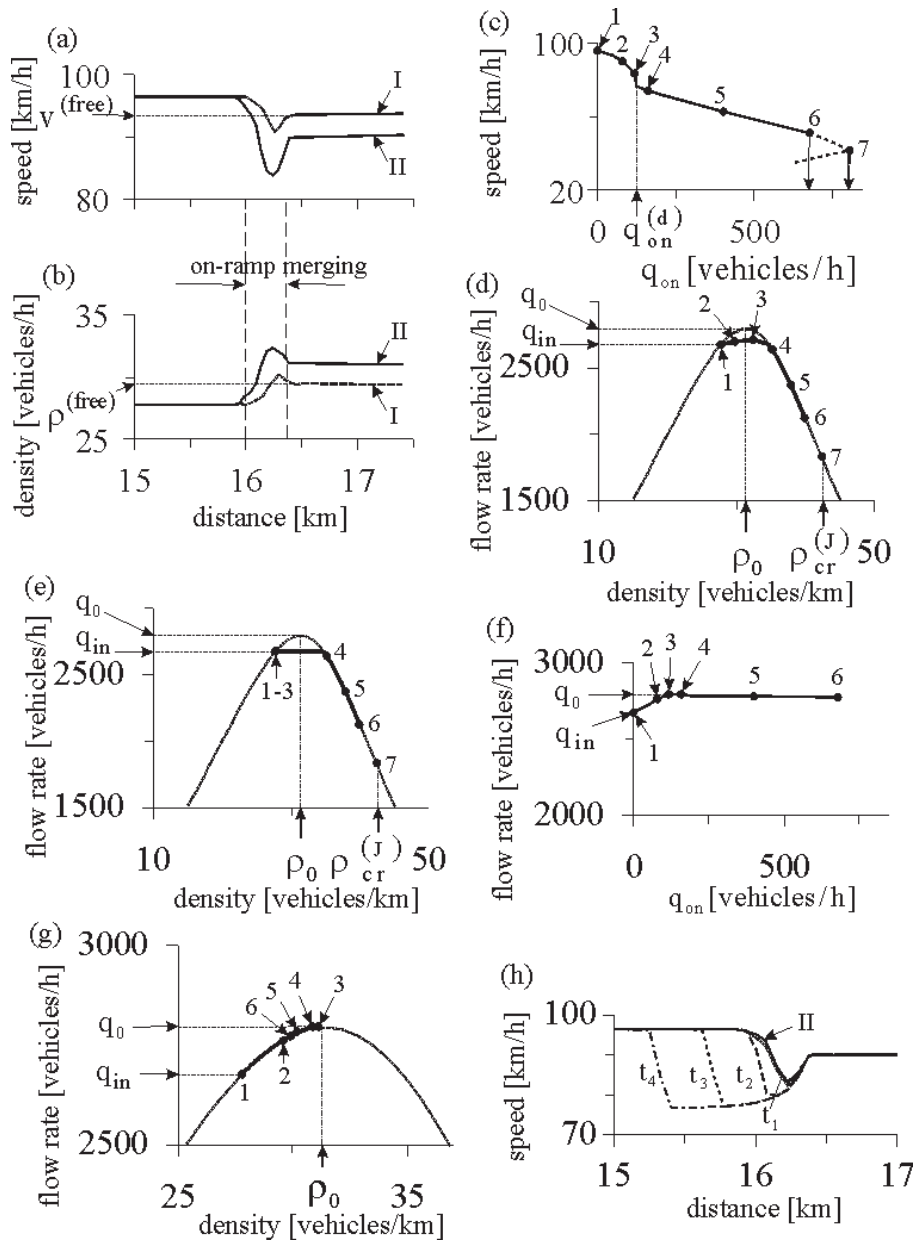}
\caption{Pattern features in the OV-model in figure~\ref{OV_patterns} (b)
at an on-ramp bottleneck:
 (a, b) -- Spatial dependences of the averaged speed  (a) and density (b)  on the main road
 within   deterministic perturbations  localized at the bottleneck
at    $q_{\rm on}=$ 80 (curve I) and 117 (curve II) vehicles/h.
(c, d) -- On-ramp flow rate dependencies of the average speed   (c), flow rate
and density (d) on the main road at   locations of the minimum of the average speed. 
 (e) -- On-ramp flow rate dependences of the  flow rate
and density on the main road at the location 200 m upstream of the begin
 of the on-ramp merging region.
 (f, g) -- On-ramp flow rate dependences of the  flow rate (f)
and density (in the flow--density plane) (g) on the main road in free flow downstream of the bottleneck. 
(h) -- Wave of dense flow that starts to propagate upstream with the velocity $v_{\rm d}\approx -$ 0.7 km/h
 under the condition (\ref{maxq0_delta})
at a very small value $\Delta q= $ 3 vehicles/h; $t_{1}=15$, $t_{2}=25$, $t_{3}=55$,
$t_{4}=90$ min; $q_{\rm on}=$ 120 vehicles/h; the deterministic perturbation (curve II) is the same as those in (a).
Dotted curves on   (d, e, g) show steady model states in the
flow--density plane.
 $q_{\rm in}=$ 2676 vehicles/h, $q_{0}=$ 2795 vehicles/h.  5-min averaged data.
\label{OV_Perturbation} } 
\end{center}
\end{figure*}

In the case (ii) (figure~\ref{OV_patterns} (b)), the flow rate in free flow downstream of an on-ramp bottleneck $q_{\rm sum}$   
cannot exceed the maximum flow rate on the fundamental diagram $q_{0}$. If  $q_{\rm in}$ is a given large
enough value and the flow rate $q_{\rm on}$ begins to increase, then a localized perturbation  as
 that in the ATD- and SA-models
(Sects.~\ref{ATD_LP_Sect} and~\ref{ATD_SA_Meta})
appears at the bottleneck
(figures~\ref{OV_Perturbation} (a) and (b)). 
The minimum speed   within the time-averaged (deterministic)
 perturbation on the main road decreases when $q_{\rm on}$ increases
(points 1--3 in   figure~\ref{OV_Perturbation} (c)).
In the OV-model, when    $q_{\rm on}$ increases, the location of the minimum speed  
within the deterministic perturbation on the main road exhibits firstly a slight shift
downstream and then upstream within the merging region of the on-ramp (figure~\ref{OV_Perturbation}
 (a, b)).  For this reason,  the flow rate on the main road at the location of the minimum speed within the deterministic perturbation on the main road
firstly slightly increases and then decreases (points 1--3 in   figure~\ref{OV_Perturbation} (d)),
whereas the flow rate on the main road upstream of the on-ramp merging region  
 is  equal to  $q_{\rm in}$ (points 1--3 in   figure~\ref{OV_Perturbation} (e)).
When $q_{\rm on}$ increases beginning from zero, the flow rate $q_{\rm sum}$ downstream of the bottleneck increases
beginning  from
$q_{\rm sum}=q_{\rm in}$ (points 1--3 in   figures~\ref{OV_Perturbation} (f) and (g)).

 At a given large enough flow rate $q_{\rm in}$
 this growth of the local perturbation in free flow at the bottleneck with $q_{\rm on}$ has a limit. This limit is reached when
 the flow rate $q_{\rm on}$ reaches some critical value
 $q_{\rm on}=q^{\rm (d)}_{\rm on}$ at which
the  flow rate $q_{\rm sum}$  is equal to the
 maximum flow rate on the fundamental diagram:  
\begin{equation}
q_{\rm sum}=q_{\rm in}+q^{\rm (d)}_{\rm on}=q_{\rm 0}. 
\label{maxq0}
\end{equation}
When the flow rate $q_{\rm on}$ increases further, i.e.,
 \begin{equation}
\Delta q=q_{\rm in}+q_{\rm on}-q_{\rm 0}>0, 
\label{maxq0_delta}
\end{equation}
then the upstream front of the initial perturbation,
which is motionless at the condition $q_{\rm sum}=q_{\rm in}+q_{\rm on}\leq q_{\rm 0}$
(curve II in figure~\ref{OV_Perturbation} (h)),
begins to move upstream of the bottleneck, i.e., a wave of lower speed and greater
density propagating upstream appears (spatial 
speed distributions related to the times $t_{1}$--$t_{4}$ in figure~\ref{OV_Perturbation} (h)).
As a result, a dense flow 
associated with the branch of the diagram with a negative slope
 occurs upstream of the bottleneck (figures~\ref{OV_patterns} (c) and (d) and points 4--6 in
 figures~\ref{OV_Perturbation} (c)--(e)).
At the critical point  (\ref{maxq0}), the   derivative of the minimum average speed on the main road
on the flow rate $q_{\rm on}$ is discontinuous, whereas this speed is a continuous decreasing function of $q_{\rm on}$
(figure~\ref{OV_Perturbation} (c)). 
The greater the flow rate $q_{\rm on}$, 
 specifically, the greater $\Delta q$ (\ref{maxq0_delta}),
the greater absolute velocity of the wave of dense flow propagation $\mid$$v_{\rm d}$$\mid$ (figures~\ref{OV_patterns} (c) and (d)).
In addition, the flow rate   downstream of the bottleneck, which is equal to $q_{\rm 0}$
under the condition (\ref{maxq0}), remains approximately  to be equal to  $q_{\rm 0}$,
when 
$q_{\rm on}$ increases (points 4--6 in   figures~\ref{OV_Perturbation} (f) and (g)).

It must be noted that   the above mentioned
 behaviour of the upstream front of the perturbation
at the bottleneck in the OV model (\ref{a_OV}), (\ref{FJ_02})
 is qualitatively different from those for the   upstream front of
 the perturbation at the bottleneck in the ATD- and SA-models. In the latter case,
 when the flow rate $q_{\rm on}$ reaches the critical value for an F$\rightarrow$S transition,
 a wave
 of synchronized flow occurs abruptly and   propagates upstream with a {\it finite}   velocity.
 This is associated with a first-order F$\rightarrow$S transition.
 In contrast, in the OV-model there is {\it no} discontinuous change in the velocity
 $v_{\rm d}$ when due to an increase in $q_{\rm on}$ the condition (\ref{maxq0_delta})
 is satisfied: $\mid$$v_{\rm d}$$\mid$ increases continuously beginning from zero, when $q_{\rm on}$ first reaches and then exceeds the critical 
flow rate $q^{\rm (d)}_{\rm on}$ associated with the condition (\ref{maxq0}). Specifically, we find that if $\Delta q\rightarrow 0$, then
 $\mid$$v_{\rm d}$$\mid$$\rightarrow 0$. Thus, in the OV-model there is no first-order
 phase transition from free flow to dense flow.

The widening dense flow upstream of the bottleneck (figures~\ref{OV_patterns} (c) and (d)
and~\ref{OV_Perturbation} (h)) can exist only, 
when the density $\rho_{\rm d}$ in the dense flow 
 satisfies
the condition
\begin{equation}
\rho_{\rm 0}<\rho_{\rm d}<\rho^{\rm (J)}_{\rm cr}. 
\end{equation}
This is  because at the density $\rho_{\rm d}=\rho^{\rm (J)}_{\rm cr}$ 
the dense flow loses its stability against wide
moving jam emergence (point 7 and
 dotted down-arrow in   figure~\ref{OV_Perturbation} (c)). However,  dynamic waves that
 emerge due to vehicle merging at the bottleneck propagate through the dense flow. For this reason,
 in numerical simulations
this moving jam emergence occurs already   at the density $\rho_{\rm d}<\rho^{\rm (J)}_{\rm cr}$
 (point 6 and
 solid down-arrow  in   figure~\ref{OV_Perturbation} (c)).

  The congested patterns in figures~\ref{OV_patterns} (d) and (e) at the first glance resemble a widening SP
  and an GP, respectively. Indeed,  in both cases a dense flow occurs upstream of the bottleneck whose
    downstream front  is fixed at the bottleneck. Thus, this dense flow should satisfy
    the macroscopic spatiotemporal objective criteria for the synchronized flow phase (Sect.~\ref{Introduction}).
    This conclusion is, however, incorrect. To explain this, note that in {\it empirical} observations
    application of the objective criteria, which define
      the traffic phases in congested traffic,  leads to clear distinction of  the synchronized flow phase.
    This    synchronized flow exhibits the following fundamental {\it empirical} feature: An F$\rightarrow$S transition
    leading to synchronized flow emergence is a first-order phase transition.
    In contrast, in a traffic flow  {\it model} an application of the objective criteria does not guarantee
    that  dense  flow occurrence in free flow is associated with
    a first-order phase transition, which is one of the requirements for the synchronized flow phase.
    
    This conclusion concerns  the OV model   (\ref{a_OV}), (\ref{FJ_02}) (figure~\ref{OV_patterns} (b)) 
     as well as   other models in the context of the fundamental diagram approach  under   condition (\ref{FJ_02}).
    Whereas for the SA-model there is a Z-shaped speed--flow characteristic
    associated with  a first-order F$\rightarrow$S transition  in
    free flow at the bottleneck (figures~\ref{SA_Pert} (c)--(e)), for the OV model the on-ramp flow rate dependence of the speed
    at the bottleneck is  a {\it monotonous}
    decreasing function (figure~\ref{OV_Perturbation} (c)): There is no first-order phase transition, when
    a dense flow related to the fundamental diagram with a negative slope
    is formed upstream of the bottleneck. Thus, the dense traffic flow in the case of the OV model and other 
     models in the context of the fundamental diagram approach  under   condition (\ref{FJ_02}) does not exhibit
     the  important empirical feature of synchronized flow and, therefore, the dense flow is not associated with
     the synchronized flow phase.

   There are also 
   traffic flow models
  in the context of the fundamental diagram approach, in which
 there is no instability of steady model states on the fundamental diagram 
 regardless of the vehicle density.
   Examples of this model class are as follows: (i) An OV model
   (\ref{a_OV}) in which the sensitivity   $A(g, \ v)$ is great enough regardless of $v$ and $g$. 
   (ii) The Nagel-Schreckenberg cellular automata model
  in the deterministic model limit, i.e., when probability of model
  fluctuations in this model is equal zero ($p=0$)~\cite{NS}. (iii) The Lighthill-Whitham-Richards  
  model~\cite{LW} and the associated  cell-transmission models~\cite{Daganzo}. In this model class,
    traffic patterns at a freeway bottleneck are qualitatively similar as those
   found in the OV model (\ref{a_OV}), (\ref{FJ_02}) at 
  the density  considerably smaller than the critical density
  $\rho^{\rm (J)}_{\rm cr}$
  (the patterns associated with points 1--5 in figures~\ref{OV_Perturbation} ({\it c})--({\it g})).
    These common 
    model features   are as follows: 
 1)  the local perturbation at the bottlenecks at $\Delta q< 0$
 (figures~\ref{OV_Perturbation} ({\it a}) and ({\it b}));  2)
 widening dense
     flow upstream of the bottleneck at $\Delta q> 0$ (figures~\ref{OV_patterns}({\it c}) and ({\it d})
     and~\ref{OV_Perturbation} ({\it h})); 
     3) there is no discontinuous change in speed (no speed breakdown) at the bottleneck when  
     widening dense
     flow occurs; 4) with an increase in
     traffic demand  at $\Delta q\geq 0$, the upstream front velocity of widening dense
     flow increases continuously beginning from zero. 
 Thus, in this model class, 
  there is no   first-order F$\rightarrow$S transition observed during  the onset of congestion at the bottleneck, i.e., this dense flow 
  has no relation to real freeway traffic.

\subsection{Conclusions \label{Conc}}

(i) Two  different deterministic microscopic traffic flow
model classes in the context of three-phase traffic theory, the ATD- and SA-models, have been introduced in the article.

(ii) The ATD- and SA-models reproduce  important {\it empirical} spatiotemporal features of phase transitions
in traffic flow 
and congested traffic patterns.

(iii) In contrast with all other known deterministic microscopic traffic flow models, in the ATD- and SA-models
vehicles moving in free flow and vehicles moving in synchronized flow exhibit
qualitatively different dynamic  behaviour. This is a result of the introduction of two separated 
regions of steady state model solutions for free flow and synchronized flow in the ATD- and SA-models
as well as  
different dynamic rules of vehicle motion in free flow and synchronized flow  implemented
in the models.

(iv) As in empirical observations, there is a first-order phase transition in the ATD- and SA-models
from free flow to synchronized flow that explained the onset of congestion at bottlenecks in these models.  

(v) The  nature of the onset of congestion as a first-order F$\rightarrow$S transition in free flow at the bottleneck, which 
the ATD- and SA-models show, is also associated with metastability of free flow at the bottleneck against
external short-time disturbances in this flow in a neighbourhood of the
bottleneck. As a result, there is multiple congested pattern emergence in an initial free flow at the bottleneck in the
 ATD- and SA-models: Depending on
an amplitude (or duration) of
an external disturbance, one of the SPs or else an GP can be induced in free flow at the bottleneck
at the same chosen model parameters.

(vi) In accordance with empirical results, in the ATD- and SA-models moving jams can emerge spontaneously in synchronized flow only, i.e.,
as a result of F$\rightarrow$S$\rightarrow$J transitions.

(vii) In addition to the above common behaviour of the ATD- and SA-models, these models
exhibit also some qualitatively different  features. This is because in the ATD-model synchronized flow model steady 
states are related to a 2D-region in the flow--density plane, whereas synchronized flow model steady 
states in the SA-models belong to an 1D-region (a curve) in the flow--density plane. In particular, the following differences of model features 
have been found:

(1) The ATD-model can show all
types of spatiotemporal congested patterns at an on-ramp bottleneck observed in empirical observations.

(2)
In contrast,   SA-models  cannot
 show LSPs associated with empirical results as well as some of empirical   features of synchronized flow between wide moving jams within
 GPs.
 
(viii) Models in the context of the fundamental diagram approach reviewed 
in~\cite{Gartner,Wolf,Sch,Helbing2001,Nagatani_R,Nagel2003A} 
cannot explain the onset of congestion in free flow, which in empirical observations is associated with
a first-order F$\rightarrow$S transition. Depending on the model type and model parameters, 
in these models either wide moving jam emergence is responsible for the onset of congestion
at an on-ramp bottleneck rather than an empirically observed F$\rightarrow$S transition or a widening dense
traffic flow  occurs upstream of the bottleneck
when the density in free flow at the bottleneck exceeds the density associated with the maximum point on the
fundamental diagram. In the latter case, in contrast with empirical observations
  there is {\it no} 
first-order phase transition from an initial free flow to this   dense   flow 
  at the bottleneck: The   dense flow results  from non-homogeneity of a freeway in a neighbourhood of
 the bottleneck. Thus, these models cannot show a first-order F$\rightarrow$S transition observed during  the onset of congestion at the bottleneck
 in real freeway traffic, i.e., this dense flow 
  has no relation to real freeway traffic. Indeed,
 the first-order F$\rightarrow$S transition is a fundamental empirical feature of the onset of congestion in free
 flow with the subsequent synchronized flow phase emergence at the bottleneck.

\appendix

\section{}
\label{A}

To derive   formula (\ref{a_safety_app})~\cite{KK2005A},
let us consider
a solution of
(\ref{Gipps_safety})
 when it is an equality:
\begin{eqnarray}
\label{v_s}
v_{\rm s} (g, \ v_{\ell})= \frac{2b_{\rm s}g+v^{2}_{\ell}}{b_{\rm s}T_{\rm s}+
\sqrt{{b^{2}_{\rm s}T^{2}_{\rm s}+2b_{\rm s}g+v^{2}_{\ell}}}}.
\end{eqnarray}
From (\ref{Gipps_safety}), (\ref{v_s}), it can be seen that if
$g=v_{\ell}T_{\rm s}$, then the safe speed
$v_{\rm s}=v_{\ell}$; if in contrast $g<v_{\ell}T_{\rm s}$, then the speed $v_{\rm s}< v_{\ell}$.
 In particular, this ensures  collision less vehicle motion.
To simplify the formula (\ref{v_s}),
let us replace the space gap $g$ in denominator of (\ref{v_s})
by the value $v_{\ell}T_{\rm s}$. This reduces the safe speed
$v_{\rm s}$ at $g < v_{\ell}T_{\rm s}$, therefore, the safety condition (\ref{Gipps_safety})  remains to be valid.
 Then from formula (\ref{v_s}), we get
\begin{eqnarray}
\label{v_s0}
v_{\rm s} (g, \ v_{\ell})= \frac {g+v^{2}_{\ell}/(2 b_{\rm s})}{T_{\rm s}+
v_{\ell}/(2 b_{\rm s})}.
\end{eqnarray}
To provide more comfortable vehicle deceleration,
an anticipated  gap $g^{\rm (a)}=g + (v_{\ell}-v) T_{0}$ is used in formula
(\ref{v_s0})
rather than the gap $g$.
As a result,  (\ref{v_s0}) takes the form
\begin{eqnarray}
\label{v_s1}
v_{\rm s} (g, \ v_{\ell})= \frac {g + (v_{\ell}-v) T_{0}+v^{2}_{\ell}/(2
b_{\rm s})}{T_{\rm s}+ v_{\ell}/(2 b_{\rm s})}.
\end{eqnarray}
Substituting  (\ref{v_s1}) into (\ref{a_safety}), we find formula
(\ref{a_safety_app})
with coefficiens (\ref{a_safety_coef}), (\ref{a_safety_coef1}).

Note that we have also tested  another formulation for the safe speed in the ATD-model when
the speed $v_{\rm s}(g, \ v_{\ell})$ in (\ref{a_safety}) is given by  formula (\ref{v_s}).
Simulations of the ATD-model show that
both formulations (\ref{a_safety_app}) and (\ref{v_s})  ensure  collision less
vehicle motion at an appropriate choice of model parameters
and   lead to qualitatively the same features of phase transitions and
congested patterns.

\section{}
\label{B}

In this Appendix, two further variants of the  SA-models are presented.
In the first of these variants,
the formula (\ref{acc_sa}) reads as follows
\begin{eqnarray}
\label{acc_sa2}
\frac{dv}{dt}=\left\{
\begin{array}{ll}
a^{\rm (free)} &  \textrm{at $g \geq g^{\rm (free)}_{\rm min}$ },  \\
a^{\rm (syn)} &  \textrm{at $ g^{\rm (jam)}_{\rm max} < g < g^{\rm (free)}_{\rm min}$}, \\
a^{\rm (jam)} &  \textrm{at $0 \leq  g \leq g^{\rm (jam)}_{\rm max}$},
\end{array} \right.
\end{eqnarray}
where $a^{\rm (free)}$, $a^{\rm (syn)}$, $a^{\rm (jam)}$ are given by (\ref{acc_lim_sa})--(\ref{v_syn_sa_2}).
In this SA-model,
steady states of free flow
(the curve $F$
in figure~\ref{Steady_states_SA} (e)) correspond to the condition  (\ref{steady_free_sa}),
averaged steady states of synchronized flow
are related to a line $S$   given by the condition
 \begin{equation}
q =  (1-\rho/ \rho^{\rm (jam)}_{\rm min})/T^{\rm (syn)}_{\rm av}
 \quad {\rm at} \  \rho^{\rm (free)}_{\rm max}< \rho<\rho^{\rm (jam)}_{\rm min},
\label{steady_syn0_sa3}
\end{equation}
  steady states for a wide moving jam
are associated with the condition
(\ref{steady_jam0_sa})  (figure~\ref{Steady_states_SA} (e)).

In another variant of SA-model, formula (\ref{acc_sa}) reads as follows
\begin{eqnarray}
\label{acc_sa3}
\frac{dv}{dt}=\left\{
\begin{array}{ll}
a^{\rm (free)} &  \textrm{at $g \geq g^{\rm (free)}_{\rm min}$},  \\
a^{\rm (FS)} &  \textrm{at $ g^{\rm (syn)}_{\rm max} < g < g^{\rm (free)}_{\rm min}$}, \\
a^{\rm (syn)} &  \textrm{at $g^{\rm (jam)}_{\rm max} < g \leq g^{\rm (syn)}_{\rm max}$}, \\
a^{\rm (jam)} &  \textrm{at $0 \leq  g \leq g^{\rm (jam)}_{\rm max}$}.
\end{array} \right.
\end{eqnarray}
  In (\ref{acc_sa3}), $g^{\rm (syn)}_{\rm max}$ is the maximum space gap in  synchronized flow;
$a^{\rm (jam)}$,   $a^{\rm (free)}$, $a^{\rm (syn)}$ are given by (\ref{acc_lim_sa}) in which 
  $\tilde a^{\rm (jam)}$ is taken from (\ref{a_jam_sa}), 
   \begin{eqnarray}
\label{a_free_sa4}
\tilde a^{\rm (free)}(g, \ v, \ v_{\ell})=
A^{\rm (free)}(V^{\rm (free)}(g)-v)+  \nonumber \\
K^{\rm (free)}(v_{\ell}-v),
\end{eqnarray}
\begin{eqnarray}
\label{a_syn_sa4}
\tilde a^{\rm (syn)}(g, \ v, \ v_{\ell})=A^{\rm (syn)}
\big(V^{\rm (syn)}_{\rm av}(g)-v\big)+ \nonumber  \\
K^{\rm (syn)}(v, \ v_{\ell})(v_{\ell}-v), 
\end{eqnarray}
where the sensitivity
 \begin{eqnarray}
\label{Ks_sa4}
K^{\rm (syn)}(v, \ v_{\ell})=\left\{
\begin{array}{ll}
K^{\rm (acc)} &  \textrm{at $v<v_{\ell}$}, \\
K^{\rm (dec)} &  \textrm{at $v\geq v_{\ell}$},
\end{array} \right.
\end{eqnarray}
 $K^{\rm (free)}$ is a sensitivity, $V^{\rm (syn)}_{\rm av}(g)$ is given by (\ref{v_syn_sa_2}).
A function $a^{\rm (FS)}(g, \ v, \ v_{\ell})$
 in (\ref{acc_sa3}) is taken as follows
 \begin{eqnarray}
\label{a_FS_sa}
a^{\rm (FS)}(g, \ v, \ v_{\ell})=
\min\big(a_{\rm max}, \
A^{\rm (FS)} (V^{\rm (FS)}(g)-v)+ \nonumber  \\
K^{\rm (free)}(v_{\ell}-v) \big),
\end{eqnarray}
where  the function $ V^{\rm (FS)}(g)=V(g)$ at $g^{\rm (syn)}_{\rm max} < g < g^{\rm (free)}_{\rm min}$,
$A^{\rm (FS)}$ is a sensitivity that in a general case can be different from the sensitivity
$A^{\rm (free)}$ in free flow.

In the   SA-model (\ref{acc_sa3})--(\ref{a_FS_sa}),
steady states of free flow
(the curve $F$
in figure~\ref{Steady_states_SA} (f)) are associated with the condition  (\ref{steady_free_sa}),
averaged steady states of synchronized flow
 are related to a line $S$ given by the condition
 \begin{equation}
q =  (1-\rho/ \rho^{\rm (jam)}_{\rm min})/T^{\rm (syn)}_{\rm av}
 \quad {\rm at} \  \rho^{\rm (syn)}_{\rm min} \leq  \rho< \rho^{\rm (jam)}_{\rm min},
\label{steady_syn0_sa4}
\end{equation}
where $\rho^{\rm (syn)}_{\rm min}=1/(g^{\rm (syn)}_{\rm max}+d)$,
and steady states for a wide moving jam
are found from the condition
(\ref{steady_jam0_sa})  (figure~\ref{Steady_states_SA} (f)).

In contrast with the other SA-models, the SA-model (\ref{acc_sa3})--(\ref{a_FS_sa})   has
a limited density range of steady states between steady states for free flow and synchronized flow,
which are found from the condition
\begin{eqnarray}
a^{\rm (FS)}=0   \quad \textrm{at $ g^{\rm (syn)}_{\rm max} < g < g^{\rm (free)}_{\rm min}$ }.
\label{a_FS0_sa3}
\end{eqnarray}
Eq. (\ref{a_FS0_sa3}) yields the following condition for these steady states   in the flow--density plane (curve $FS$
in figure~\ref{Steady_states_SA} (f))
\begin{eqnarray}
q =\rho V_{\rm F}(\rho)  \quad {\rm at} \ \rho^{\rm (free)}_{\rm max} < \rho< \rho^{\rm (syn)}_{\rm min},
\label{steady_FS0_sa3}
\end{eqnarray}
where the density $\rho^{\rm (free)}_{\rm max}$ at the maximum point for free flow is not greater than
the density $\rho_{0}$ associated with the maximum point on the curve $FS$ (figure~\ref{Steady_states_SA} (f)).
To simulate a first-order  F$\rightarrow$S   transition, 
the steady state model solutions (\ref{steady_FS0_sa3}) should be {\it unstable} against infinitesimal
non-homogeneous perturbations. This requirement to the SA-model (\ref{acc_sa3})--(\ref{a_FS_sa})
is easy satisfied
  through an appropriated choice of the function $V^{\rm (FS)}(g)$ and
the sensitivities $A^{\rm (FS)}$, $K^{\rm (free)}$,  $K^{\rm (syn)}$. 
In this case, numerical simulations of
the SA-model (\ref{acc_sa3})--(\ref{a_FS_sa}) made  show that this model 
exhibits F$\rightarrow$S$\rightarrow$J transitions in accordance with
empirical results (we used the following parameters
for the SA-model (\ref{acc_sa3})--(\ref{a_FS_sa}): $V(g)=V_{0} \big(\tanh((g-g_{0})/g_{1})+\tanh(g_{0}/g_{1}) \big)$ 
with $V_{0}=$ 14 m/s,
 $g_{0}=$ 21 m,  $g_{1}=$ 7 m; $g^{\rm (syn)}_{\rm max}=$ 24 m; $A^{\rm (free)}=A^{\rm (FS)}= \ 0.1 \ s^{-1}$;
$K^{\rm (free)}= \ 0.6 \ s^{-1}$;  $K^{\rm (acc)}= \ 0.4 \ s^{-1}$; $K^{\rm (dec)}$ is taken from Table~\ref{tableATD1} with 
$K^{\rm (dec)}_{1}=\ 1 \ s^{-1}$, $v_{\rm c}=$ 9 m/s, $\epsilon=$ 0.05; other parameters are the same as those in
 the SA-model 
(\ref{coor_sa})--(\ref{v_syn_sa_2})).

\end{document}